\documentclass[numberedappendix]{emulateapj}
\usepackage{graphics}
\newcommand{\oii}{\mbox{[O~{\sc ii}]} \mbox{$\lambda \lambda 3728$}}
\newcommand{\oiiia}{\mbox{[O~{\sc iii}]} \mbox{$\lambda 4960$}}
\newcommand{\oiiib}{\mbox{[O~{\sc iii}]} \mbox{$\lambda 5008$}}
\newcommand{\ha}{H$\alpha$ $\lambda 6565$}
\newcommand{\hb}{H$\beta$ $\lambda 4863$}
\newcommand{\hg}{H$\gamma$ $\lambda 4342$}
\newcommand{\hd}{H$\delta$ $\lambda 4103$}
\newcommand{\niia}{\mbox{[N~{\sc ii}]} $\lambda 6550$}
\newcommand{\niib}{\mbox{[N~{\sc ii}]} $\lambda 6585$}
\newcommand{\siia}{\mbox{[S~{\sc ii}]} $\lambda 6718$}
\newcommand{\siib}{\mbox{[S~{\sc ii}]} $\lambda 6733$}

\shorttitle{SDSS Spectro Lens Search. I.}
\shortauthors{Bolton et al.}

\begin{document}
\journalinfo{To appear in The Astronomical Journal, 2004 April}
\submitted{AJ, in press, 2004 April}

\title{Sloan Digital Sky Survey Spectroscopic
       Lens Search. \\ I. Discovery of Intermediate-Redshift
       Star-Forming \\ Galaxies Behind Foreground Luminous
       Red Galaxies\altaffilmark{1}}

\author{Adam~S.~Bolton\altaffilmark{2},
        Scott~Burles\altaffilmark{2},
        David~J.~Schlegel\altaffilmark{3},
        Daniel~J.~Eisenstein\altaffilmark{4},
        and J.~Brinkmann\altaffilmark{5}}
\altaffiltext{1}{Based in part on observations
       obtained with the 6.5-m Clay telescope of the Magellan
       Consortium.}
\altaffiltext{2}{Department of Physics and Center for Space Research,
    Massachusetts Institute of Technology,
    77 \mbox{Massachusetts} Avenue, Cambridge, MA 02139
    ~{\tt (bolton@mit.edu, burles@mit.edu)}}
\altaffiltext{3}{Princeton University Observatory, Princeton, NJ
    08544-1001 ~{\tt (schlegel@astro.princeton.edu)}}
\altaffiltext{4}{Steward Observatory, University of Arizona,
    933 North Cherry Avenue, Tucson, AZ 85721}
\altaffiltext{5}{Apache Point Observatory, P.O. Box 59,
    Sunspot, NM 88349}

\begin{abstract}
We present a catalog of 49 spectroscopic strong gravitational
lens candidates selected from a Sloan Digital Sky Survey
sample of 50996 luminous red galaxies.
Potentially lensed star-forming galaxies
are detected through the presence of
background oxygen and hydrogen nebular
emission lines in the spectra of these massive
foreground galaxies.  This multiline selection eliminates
the ambiguity of single-line identification and provides a
very promising sample of candidate galaxy-galaxy lens
systems at low to intermediate redshift,
with foreground redshifts ranging from
0.16 to 0.49 and background redshifts from
0.25 to 0.81.
Any lenses confirmed within our sample
would be important new probes of
early-type galaxy mass distributions,
providing complementary constraints
to those obtained from currently known lensed
high-redshift quasars.
\end{abstract}


\section{Introduction and Motivation}

In its strongest form, gravitational lensing produces
unmistakably distorted, amplified, and multiplied images
of distant astronomical objects.  It is therefore not
surprising that the majority of known
galaxy-scale gravitational lens systems have
been discovered through imaging observations.
However, a small number of lenses have been
discovered spectroscopically, with the spectrum
of a targeted galaxy showing evidence of
emission from a background source and follow-up imaging
revealing lensing morphology.  The three most secure
examples are the lensed quasars 2237+0305 \citep{huch85}
and SDSS J0903+5028 \citep{john03} and the lensed
Lyman-$\alpha$ emitting galaxy 0047$-$2808
\citep{w96, w98, w99}.  Several authors have made
predictions for the frequency of lensed quasar discoveries
in galaxy redshift surveys \citep{koc92, mw00, mw01}.
Others, inspired by the
discovery of 0047$-$2808,
have undertaken spectroscopic
searches for lensed Lyman-$\alpha$-bright galaxies
\citep{h00, wil00, hall00, behs}.
The idea behind such searches
is that a massive foreground galaxy should act as
an effective gravitational lens of {\em any}
object positioned
sufficiently far behind it and at small enough
impact parameter, and any emission features
from such lensed objects should be detectable
in the spectra of
the foreground galaxy.  Therefore a search for
discrepant emission features in galaxy spectra
can potentially
lead to a sample of gravitational lens systems that
would not be discovered in broadband imaging searches
due to faintness of source relative to lens.
This ``lenses-looking-for-sources'' approach
is complementary to lens searches such as
the recently completed Cosmic Lens All-Sky Survey
\citep[CLASS;][]{class1, class2} which
proceed by targeting sources and looking for evidence
of an intervening lens.

With its massive scale and quality of data,
the spectroscopic component
of the Sloan Digital Sky Survey (SDSS) provides
an unprecedented opportunity for spectroscopic
galaxy-galaxy gravitational lens discovery.
This paper presents the first results of such a search
within a sample of
$\sim$51,000 SDSS Luminous Red Galaxy (LRG)
spectra \citep[hereafter E01]{lrg}:
a catalog of candidate lensed star-forming galaxies at
intermediate redshift.
The lens candidates presented in this
paper were detected by the presence of not one but
(at least) three emission lines in the LRG
spectra identified
as nebular emission from a single background redshift:
\oii\ and two out of
the three of \hb, \oiiia, and \oiiib\
(all emission-line wavelengths in this paper are
vacuum values, in Angstroms).  This implies
a maximum redshift
of $z \sim 0.8$ for any candidate lensed
galaxies: at higher redshifts, [O~{\sc iii}] emission moves
off the red end of the SDSS spectrograph.

The lensing cross section of a particular foreground
galaxy is lower for intermediate-redshift sources
than high-redshift sources, and in this sense the lens
search we describe here is at a disadvantage relative
to searches for lensed high-redshift Lyman-$\alpha$
emitters.  However, the identities of the
emission lines in the sample we present here are
absolutely unambiguous.  This cannot be said for
spectroscopic Lyman-$\alpha$ lens candidates, which are
typically detected as single discrepant emission lines
and are difficult to distinguish from lower-redshift
emission or (in the case of a huge survey such
as the SDSS) non-astrophysical spectral artifacts.
(A forthcoming paper will present SDSS single-line
[O~{\sc ii}] and Lyman-$\alpha$ lens candidates.)
Our spectroscopic lens search
based on multiple-line detection has further advantages.
First, the source redshift of any lensed galaxies will be
known from the outset.  Second, for a given limiting
line flux, star-forming
galaxies at intermediate redshift are
more numerous on the sky
than Lyman-$\alpha$ emitters
at high redshift \citep[see][]{hip03, mai03}, and hence
strong lensing events could be more frequent despite
decreased lensing cross sections.
The non-lensed source population should thus
be more amenable to study and characterization
\citep[for example]{hogg98}, facilitating lens statistical
analysis.  Finally, intermediate-redshift
lensed galaxies would probe the
mass distribution of the lens population in a
systematically different manner than do
high-redshift sources.

The search for new lens systems
is motivated by the unique power of
gravitational lensing to
constrain the mass distribution
of the lensing galaxy.
\citet{ml92} argued that strong galaxy-galaxy gravitational
lenses in the optical band
should be both ubiquitous and of great scientific
interest, but their detection has proved
difficult.  In addition to the optical Einstein ring
0047$-$2808 mentioned above,
\citet{h00} have published one more spectroscopic
galaxy-lens candidate.
{\sl Hubble Space Telescope} ({\sl HST})
imaging has also had some success in detecting
lensed galaxies, as described by \citet{cr02},
\citet{rgo99}, and \citet{fas03},
but spectroscopic confirmation
remains a challenge in most cases.
Lensed galaxies are especially useful
for investigating
the mass distribution of the lensing
galaxy because unlike lensed quasars they are
spatially resolved sources which provide more
detailed constraints \citep[for example]{kn92}.
As with lensed quasars, the relative angular
deflections between multiple images can
constrain the total deflecting lens mass.
But the relative distortions of multiple images
can provide constraints as well,
probing the slope of the radial mass
profile and allowing more detailed connection
with theories of galaxy formation
and Cold Dark Matter (CDM).
The promise of our candidate sample is
enhanced by the fact that
lens {\em and} source redshifts
for any confirmed lens systems
will be known from the outset with
great precision; this is generally not the case
for lenses discovered in optical imaging
or radio data.

The lack of sufficient lensing constraints on the
radial slope of the galaxy mass profile
in lensed quasar systems
is the greatest source of uncertainty in the
Hubble constant ($H_0$) values derived from
the time delays between images \citep{sch00},
which remain systematically low compared to
the current best $H_0$ values from the distance
ladder \citep{fre01} and cosmic microwave
background \citep{spe03};
\citet{koc02a, koc02b, koc03} highlights
the importance of resolving this
discrepancy.
Any galaxy-galaxy lenses confirmed within
our current list of candidates
would help to solve this problem
by expanding the sample for
ensemble lensing studies of early-type galaxy
mass profiles like that of \citet{rus03}
and allowing for the inclusion of distortion
information as mentioned above.
The generally lower lens
and source redshifts of our candidate systems
as compared to known lens systems are
also of particular interest because
the images of any strong lenses within our sample will
tend to form at smaller {\em physical}
radii within the lens galaxy than in current
lens systems \citep[see][Table 1]{lensevol},
thus probing elliptical galaxy mass profiles
in a systematically different
manner\footnote{The lowest-redshift lensed quasar
currently known, RXJ 1131$-$1231, has
$z = 0.658$ \citep{slu03}.}.
A large enough sample of strongly lensed star-forming
galaxies could even be monitored directly for
supernovae.  Multiple images of the same supernova
would give time delays and $H_0$ measurements,
as numerous authors have envisioned
\citep{ref64, pm00, holz01, goo02, ok03}.
If we assume one supernova per
galaxy per century as an order-of-magnitude
estimate \citep{db85}, a program to monitor
$\sim 50$ lensed galaxies for several years
would have a good chance of success.

Numerous studies have addressed the
inability of standard macroscopic galaxy
mass models to predict the flux ratios observed
between images of strongly lensed quasars.
Some authors have suggested that this
phenomenon may be due to the effects of
substructure in the lensing galaxy
on the scale of either
CDM sub-halos \citep{ms98, mm01, chi02, dk02}
or stars \citep{wms95, sw02}, while others
have invoked macroscopically
modified lens models as a possible explanation
\citep{ew03, mhb03}.
A sizeable sample of strongly
lensed galaxies would provide
valuable insight into this problem:
by virtue of the arcsecond-scale intrinsic
angular size of the source,
galaxy-galaxy lenses will be immune to the lensing effects of
any substructure with lensing deflection angles
on the scale of microarcseconds (stars) or milliarcseconds
(sub-halos).  If the suppressed saddle point image fluxes
seen in many quasar lenses are indeed due to macroscopic
inadequacies in the standard lens models,
similar effects should be seen
in galaxy-galaxy lenses.  If on the other
hand these effects are due
to micro/millilensing by substructure, they should
be absent from lensed galaxy systems.

\section{Search Sample}

The Sloan Digital Sky Survey
is a project to image roughly one-quarter of the sky
in five optical bands
and obtain spectroscopic follow-up observations
of $\sim 10^6$ galaxies and $10^5$ quasars.
\citet{sdss} provide a technical summary of the survey,
\citet{gun98} describe the SDSS camera,
\citet{fuk96}, \citet{hea01}, and \citet{sdss2jc}
discuss the photometric system and calibration,
\citet{pie03} discuss SDSS astrometry,
\citet{bla03} present the spectroscopic
plate tiling algorithm,
and \citet{edr} and \citet{dr1} describe
the survey data products.
Approximately 12\% of the galaxy spectroscopic
fibers are allocated to the
LRG sample (E01), selected to consist
of very luminous ($\ga 3 L_{\star}$)---and
hence massive---early-type galaxies at
higher redshift than the galaxies
of the main sample \citep{str02}.  We expect these
LRGs to be particularly effective gravitational lenses
of any objects positioned suitably behind them,
and we concentrate
our spectroscopic lens search on them.
We also note that LRGs should have little
dust and therefore any lensed background galaxies
should suffer minimal extinction.
The sample for our current
study consists of 50996 spectra taken between
5 March 2000 and 27 May 2003 of SDSS imaging objects
flagged as {\tt GALAXY\_RED} by the
photometric pipeline \citep{lup01}
for passing either of the
LRG cuts described by E01,
reduced by the SDSS spectroscopic
pipeline (J. Frieman et al., in preparation),
and selected to have redshifts between
0.15 and 0.65 as determined by the {\tt specBS}
redshift-finding software
(D. J. Schlegel et al., in preparation).
The low-redshift
cutoff is needed because less massive
galaxies start to pass the photometric cuts
below $z \la 0.15$
and pollute the volume-limited LRG sample;
see the discussion in E01 and the bimodal
LRG sample redshift histogram in
\citet[Fig. 14]{edr}.

\begin{figure}
\scalebox{1.15}{\plotone{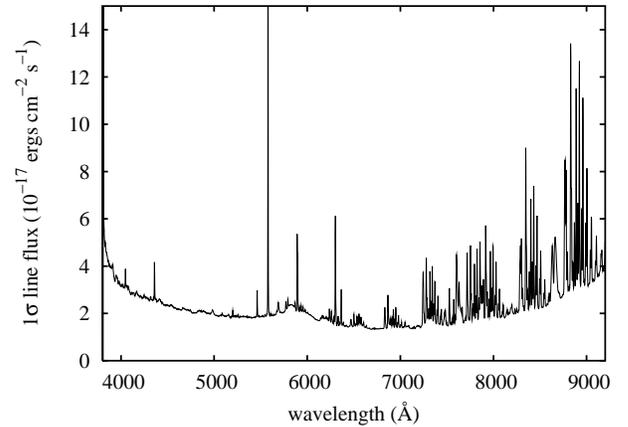}}
\caption{LRG sample median line-flux noise spectrum.
Shown is the 1-$\sigma$ noise on best-fit line fluxes
for optimal extraction of Gaussian-shaped residual emission
features with spectral width $\sigma=$1.2-pixel
($\simeq 83$ km s$^{-1}$).
Reported pixel flux variances have been rescaled as
described in Appendix~\ref{noisemodel} prior to the
calculation of this noise spectrum.
\label{noise_spec}}
\end{figure}

In addition to being much more massive than the
average galaxy, LRGs have another property
that makes them well suited to a spectroscopic
lens survey: their spectra are extremely
regular and well-characterized
\citep[see][]{eis03}.
To determine the spectroscopic redshift
of an SDSS target galaxy
with observed specific flux $f_{\lambda}$ and
one-sigma sky$+$source
noise spectrum $\sigma_{\lambda}$,
the {\tt specBS} program
employs a small set of galaxy eigenspectra
(four in the reductions for this study)
derived from a rest-frame
principal-component analysis
(PCA) of 480 galaxy spectra taken on
SDSS plate 306, MJD 51690.
This eigenbasis is incrementally redshifted,
and a model spectrum is generated from
the best-fit linear combination to
the observed spectrum at
each trial redshift, with the final redshift
assignment given by the trial value that
yields the overall minimum $\chi^2$.
Although redshift is the primary output
of this procedure,
a byproduct of {\tt specBS} is the best-fit
model spectrum itself,
$\widetilde{f}_{\lambda}$.
In the case of LRGs, $\widetilde{f}_{\lambda}$
typically provides a
very detailed and accurate fit to $f_{\lambda}$,
with a reduced $\chi^2$ of order unity over almost
4000 spectral pixels (roughly 1600 spectral
resolution elements)
attained with only 8 free parameters:
a redshift,
the four eigen-galaxy coefficients, and the three terms
of a quadratic polynomial to fit out
spectrophotometric errors and extinction effects, both
of which exist at the few-percent
level\footnote{The {\tt specBS} outputs for
the SDSS-DR1 sample are available from the
website {\tt http://spectro.princeton.edu}.}.
This extremely regular spectral behavior
allows us to form residual LRG spectra
\begin{equation}
f^{(r)}_{\lambda} \equiv f_{\lambda} - \widetilde{f}_{\lambda}
\end{equation}
that are in principle realizations of $\sigma_{\lambda}$.
Nebular emission lines from galaxies
along the line of sight other than the
target LRG will not be modeled
by {\tt specBS} and should appear as significant features
localized in wavelength within these residual spectra.
Figure~\ref{noise_spec} shows the median 1-$\sigma$
line flux sensitivity within our LRG residual spectrum
sample as a function of wavelength.  The
(20th, 50th, 80th)-percentile LRG spectra
themselves have a median signal-to-noise per pixel
of (3.3, 5.1, 9.6) at the SDSS resolution of
$\lambda / \Delta \lambda \approx 1800$.

\section{Candidate Selection}

This section describes our candidate selection routine
in detail.  Briefly stated, we select as initial candidates
those spectra that show {\em both}
blended \oii\ at $S/N > 3$
{\em and} two out of the three lines \hb, \oiiia,
and \oiiib\ at $S/N > 2.5$, then cull the candidate list
by applying cuts based on more detailed fits to the
presumed emission features, and finally remove
any obviously spurious detections.
This selection process
yields a substantial number of promising systems
without an excess of obvious false positives.

\subsection{Initial Emission Feature Detection}
\label{initial}

The key element in the first step of our
lens candidate selection
(described fully in \S~\ref{select} below)
is a straightforward matched-filtering procedure
to search for significant emission features
by fitting a Gaussian line profile
at each point in the residual spectrum
%explicitly accounting for non-stationary
%(i.e.\ wavelength-dependent) noise
\citep{pratt, h85}.
We describe our implementation here so as to
be explicit.
Let $f^{(r)}_j$ be the residual flux
in pixel $j$ and $\sigma_j^2$ be the statistical variance
of $f^{(r)}_j$ \footnote{
Conversion from units of ergs cm$^{-2}$ s$^{-1}$ \AA$^{-1}$
to units of ergs cm$^{-2}$ s$^{-1}$ pixel$^{-1}$
is made using the re-binned
SDSS spectroscopic pixel scale relation
$d \lambda = \lambda \times 10^{-4} \ln (10) \,
d(\mathrm{pixels})$.}.  Also let $\{u_i\}$ describe a
Gaussian kernel, centered
on $i = 0$, with $i$ running from $-i_{\mathrm{lim}}$
to $i_{\mathrm{lim}}$, and normalized such that
$\sum_i u_i = 1$.  The maximum-likelihood estimator
$\bar{A}_j$ for the line flux $A_j$ of any $\{u_i\}$-shaped
residual emission feature centered on pixel $j$ is that which
minimizes
\begin{equation}
\label{chi2}
\chi^2_j =
\sum_{i=-i_{\mathrm{lim}}}^{i_{\mathrm{lim}}}
\left. \left( A_j {u_i} - f^{(r)}_{(j+i)} \right)^2
\right/ \sigma_{(j+i)}^2 ~~.
\end{equation}
Differentiating (\ref{chi2}) with respect to $A_j$,
setting the resulting expression to zero, and solving yields
\begin{equation}
\bar{A}_j = C^{(1)}_j \left/ C^{(2)}_j \right. ~~,
\end{equation}
where we have defined the convolutions
\begin{eqnarray}
C^{(1)}_j &\equiv& \sum_i f^{(r)}_{(j+i)} u_i / \sigma_{(j+i)}^2 ~~,\\
C^{(2)}_j &\equiv& \sum_i u_i^2 / \sigma_{(j+i)}^2 ~~.
\end{eqnarray}
The variance of $\bar{A}_j$ (under the assumption of uncorrelated
Gaussian errors in $f^{(r)}_j$ as described by $\sigma_j^2$)
is given by
\begin{equation}
\sigma_{\bar{A},j}^2 = 1 \left/ C^{(2)}_j \right. ~~.
\end{equation}
The signal-to-noise ratio for a fitted
Gaussian profile centered on pixel $j$ is therefore
\begin{equation}
(S/N)_j = C^{(1)}_j \left/ \sqrt{C^{(2)}_j} \right. ~~.
\end{equation}
Our null hypothesis is an absence of
emission features in the residual spectra
that should manifest as the $\{(S/N)_j\}$ being
Gaussian-distributed about zero
with unit variance: this should
hold at most wavelengths in most spectra.
We approach the initial search for emission lines
in the residual spectra
as a search for significance peaks with
$(S/N)$ greater than some threshold value.
Although insensitive to goodness-of-fit, this
convolution-based detection scheme
executes quickly
and is therefore well suited to the initial search
for residual emission features within
our large spectral sample.
Appendix~\ref{noisemodel} describes a noise-rescaling
process that we employ to control the incidence of
false-positive emission-feature detections
(due primarily to imperfect sky subtraction)
without masking regions of the spectrum.

\subsection{Multi-Line Background Systems:
    Detection, Fitting, and Rejection}
\label{select}

Multiple emission features at the same
redshift will have redshift-independent wavelength
ratios.  The fully reduced SDSS spectra
have been re-binned at
a constant-velocity pixel scale of 69 km s$^{-1}$,
giving
a redshift-independent pixel offset between
features.  Our operational scheme is thus
to search for coincident
$(S/N)$ peaks between multiple copies
of a single filtered residual spectrum
that have been shifted relative to one another.
For \oii\ detection, we filter each residual spectrum
with a $\sigma=2.4$-pixel Gaussian kernel
(matched to the typical width of blended
\oii\ emission seen in SDSS starburst galaxies).
We take copies of the same residual
spectrum ``blueshifted'' by integer pixel amounts
so as to place \hd, \hg, \hb, \oiiia, \oiiib,
\niia, \ha, \niib, \siia, and \siib\ as close as
possible to 3728.48 \AA,
the geometric-mean wavelength of the \oii\ doublet.
These shifted spectra are filtered
with a $\sigma=1.2$-pixel Gaussian kernel
(matched to the typical width of SDSS
starburst \oiiib\ emission), with the sub-pixel
part of the line offset relative
to 3728.48 \AA\ incorporated by offsetting the kernel.
Any pixel in the filtered $S/N$ spectra with
value greater than 3 for \oii\ and value greater
than 2.5 for two out of \hb, \oiiia, and \oiiib\
is tagged as a ``hit''.  A group of adjacent ``hit''
pixels is reduced to the single pixel
with the greatest quadrature-sum $S/N$
for lines detected above the threshold
(in effect, the pixel most inconsistent with
the null hypothesis).
Spectra with more than one isolated hit are rejected.
The spectra are only searched in regions that would
correspond to emission from $\ga 5000$ km s$^{-1}$
behind the targeted LRG.

The preceding selection leads to 163 hits  % 142 + 21 !
within our 51,000 spectra.
For each hit, we explore a grid of 
redshift and intrinsic emission-line-width
values for the background galaxy to find a best-fit model.
At each grid point we fit a Gaussian profile
to any emission line initially
detected above a 2.0-$S/N$ threshold,
with the line center determined by the trial redshift
and line-width given by the quadrature sum
of the trial intrinsic
line-width and the measured spectrograph
resolution at the observed wavelength of each line.
\oii\ is fit with a double Gaussian profile.
We adopt as best values for
background redshift $z_{BG}$ and
(Gaussian-$\sigma$) intrinsic
line-width $\sigma_e$ those that give the minimum
$\chi^2$ over all detected lines.  The $z_{BG}$ extent
of our grid corresponds to $\pm$2 pixels, and the
explored $\sigma_e$ range runs from 0 to 2 pixels
(0 to 138 km s$^{-1}$).

Following these fits, we subject the candidate sample
to several cuts.  First, we reject any system where
no convergent (minimum $\chi^2$) value for $z_{BG}$
is found within the explored $\pm$2-pixel range:
real background emission systems should have
convergent best-fit redshifts roughly coincident with
their convolution-based $S/N$ peaks.  Similarly,
we cut systems with no convergent $\sigma_e$ between
0 and 2 pixels: the features of real systems
should neither be narrower than the spectrograph
resolution nor wider than a reasonable upper limit
for emission features of faint star-forming galaxies.
Finally, we compute a total
signal-to-noise ratio for the fit, defined as the total
best-fit flux in \oii\ and all other lines
initially detected at
$S/N \ge 2.5$ divided by the quadrature-sum of
the 1-$\sigma$ noise from those line fits,
and impose a cut in the
total-$S/N$--$\chi_r^2$ plane ($\chi_r^2$
being the $\chi^2$ per degree of freedom in the fit).
We cut any system with a total $S/N$ less than
the greater of 6 and $6 + 3(\chi_r^2 - 1)$.
This removes both low-$S/N$ candidates and candidates
whose $\chi_r^2$ values are too high to be explained
by high-$S/N$ emission features showing significant
non-Gaussian structure.
(These cuts were designed to be a
quantitative expression of our
own judgements about which candidate systems
look real upon spectrum inspection
and which do not; the
justifications are somewhat {\it a posteriori}.)
This cutting procedure reduces the 163
hits to 61 candidate systems.
Finally, we prune 12 candidates from
the list that survive
the automated culling but are clearly
explained by either
over-fit LRG stellar absorption,
under-modeled LRG line emission,
exceptionally poor data quality, or a generally
flawed template fit, leaving the 49
good candidate systems that we present.

This search for background galaxy emission lines
digs rather deep into the noise of our
spectroscopic sample.  To gauge the incidence
of false positives in our final candidate list, we
make a parallel run of
the detection, fitting, and automated rejection procedure
with the following rest-wavelength perturbations:
\hb\ $\rightarrow$ 4833, \oiiia\ $\rightarrow$ 4945,
and \oiiib\ $\rightarrow$ 5023.  These perturbations
alter all of the redshift-independent wavelength
ratios among these lines and between all of them
and \oii;  this modified detection procedure no
longer selects for real multi-line emission, but
only for noise features.  The ``false candidates''
that result from this perturbed procedure are
randomly shuffled along with the candidates
from the original procedure, and all are examined
together when making the final pruning judgements.
The perturbed procedure yields 88 hits and 7 post-cut
candidates; all 7 are pruned upon inspection without
knowledge of their intrinsic falseness.
This implies that the vast majority of our
candidates are indeed background galaxies
and not simply noise features.

\section{Candidate Systems}

\subsection{Catalog}

\begin{figure*}[h]
\scalebox{1.15}{\plotone{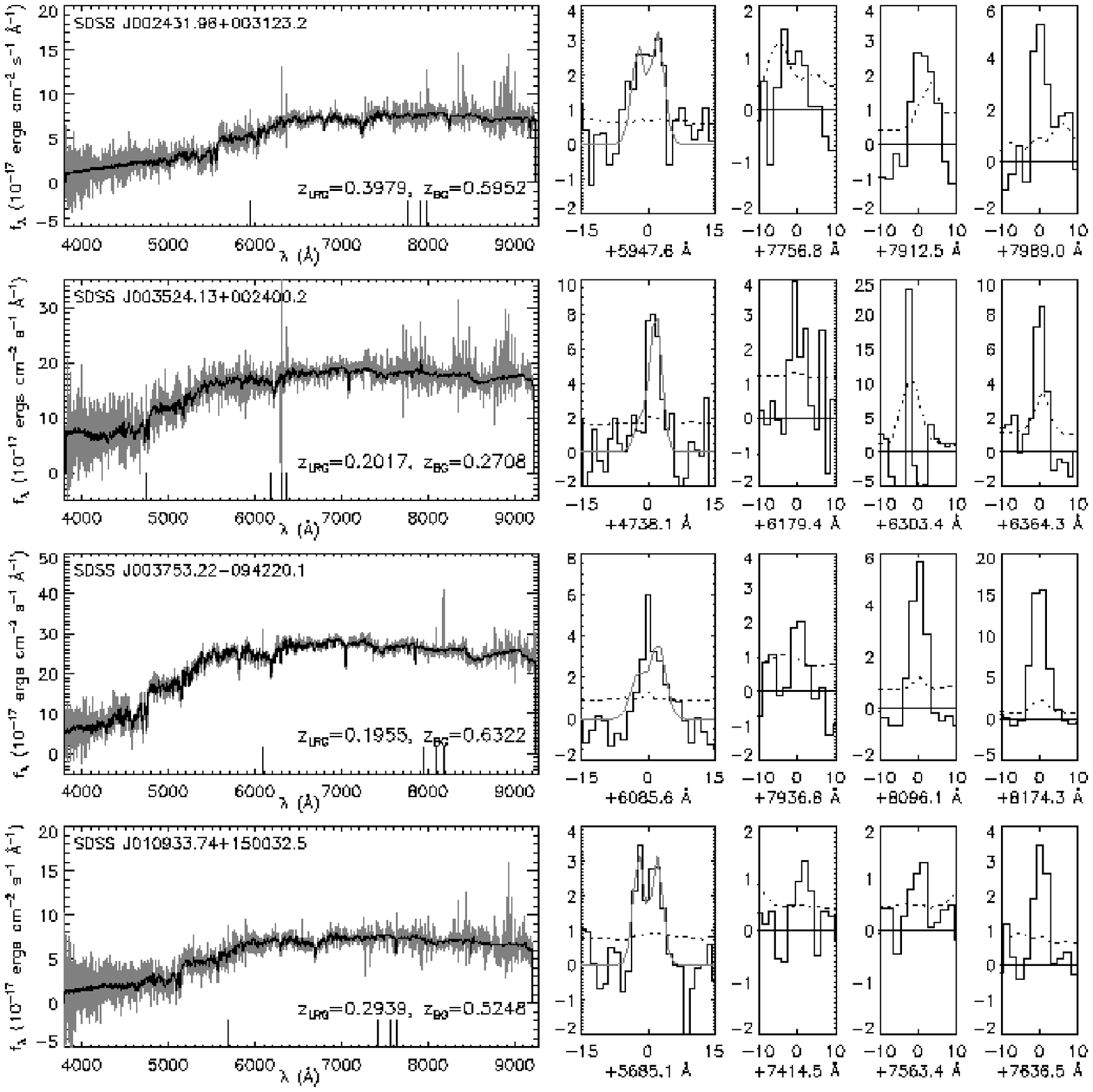}}
%\scalebox{1.15}{\plotone{bolton.fig2a.eps}}
\caption{\small SDSS LRG spectra with confirmed
background-galaxy emission features.
In the full-spectrum plots,
gray lines show data, black lines show best-fit
model spectrum, and long ticks along bottom margin
indicate the position of redshifted \oii, \hb,
\oiiia, and \oiiib\ background emission.  Smaller
windows show zooms of the residual (data $-$ model)
spectra at the positions of these same redshifted
emission lines.  Dashed lines in the smaller windows
show the 1-$\sigma$ noise level, rescaled as described
in Appendix~\ref{noisemodel}.  Gray lines in the \oii\
zoom show the double-Gaussian fit; the zero-flux line is
drawn in the other three zooms.
The spectroscopic resolution is
$\lambda / \Delta \lambda \approx 1800$.
Note the changing
vertical scales.
\label{spectra}}
\end{figure*}

\addtocounter{figure}{-1}
\begin{figure*}
\scalebox{1.15}{\plotone{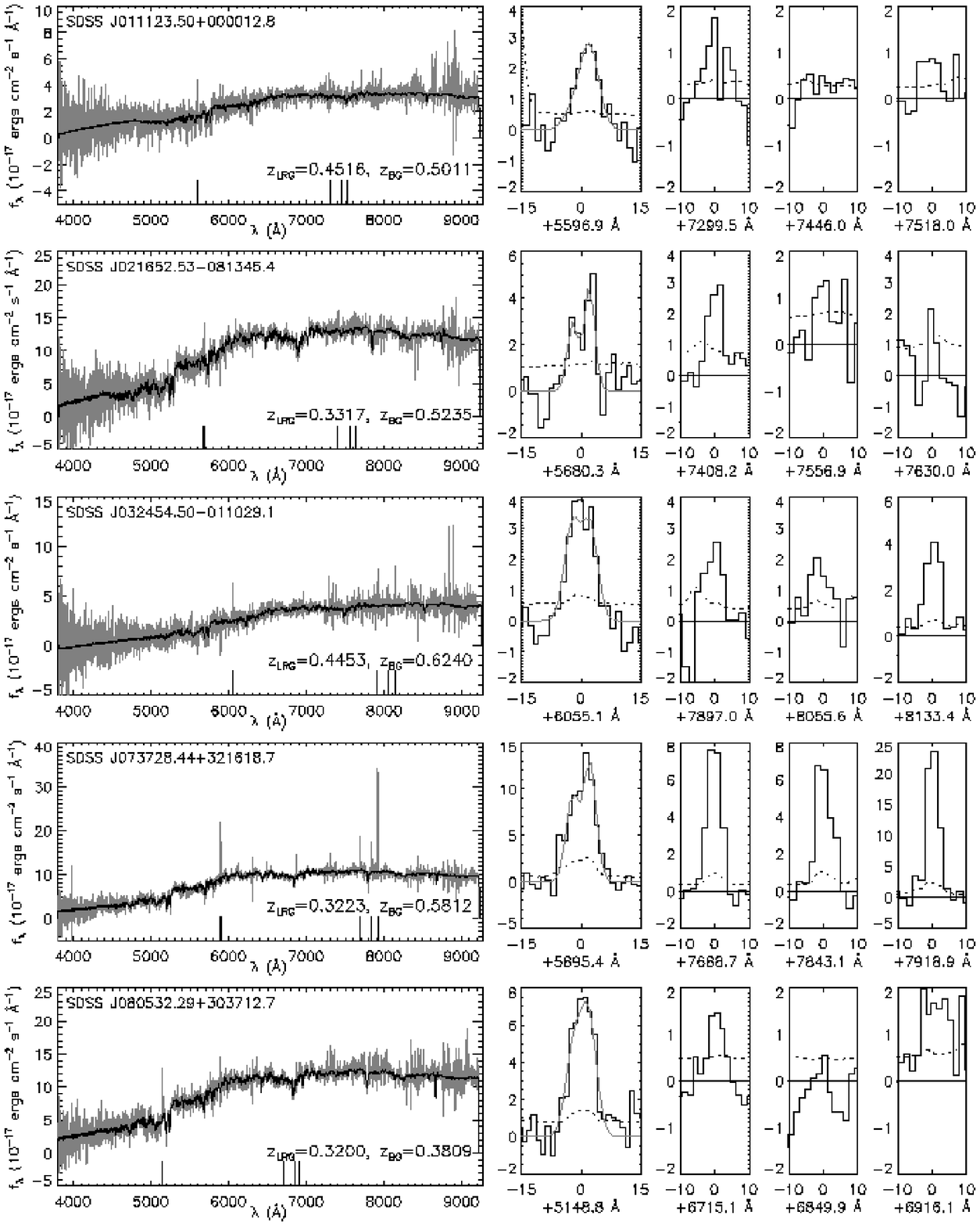}}
%\scalebox{1.15}{\plotone{bolton.fig2b.eps}}
\caption{Continued}
\end{figure*}

\addtocounter{figure}{-1}
\begin{figure*}
\scalebox{1.15}{\plotone{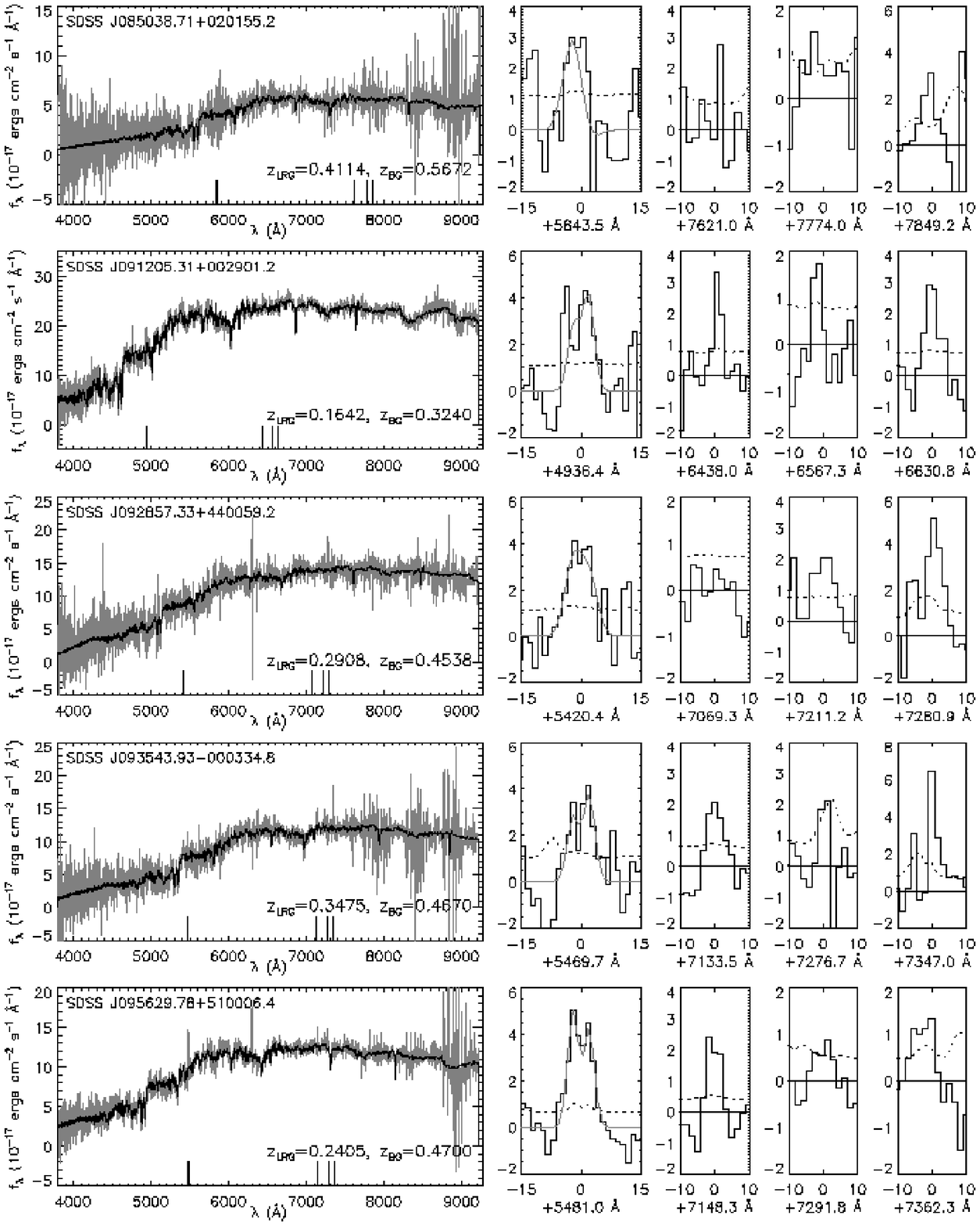}}
%\scalebox{1.15}{\plotone{bolton.fig2c.eps}}
\caption{Continued}
\end{figure*}

\addtocounter{figure}{-1}
\begin{figure*}
\scalebox{1.15}{\plotone{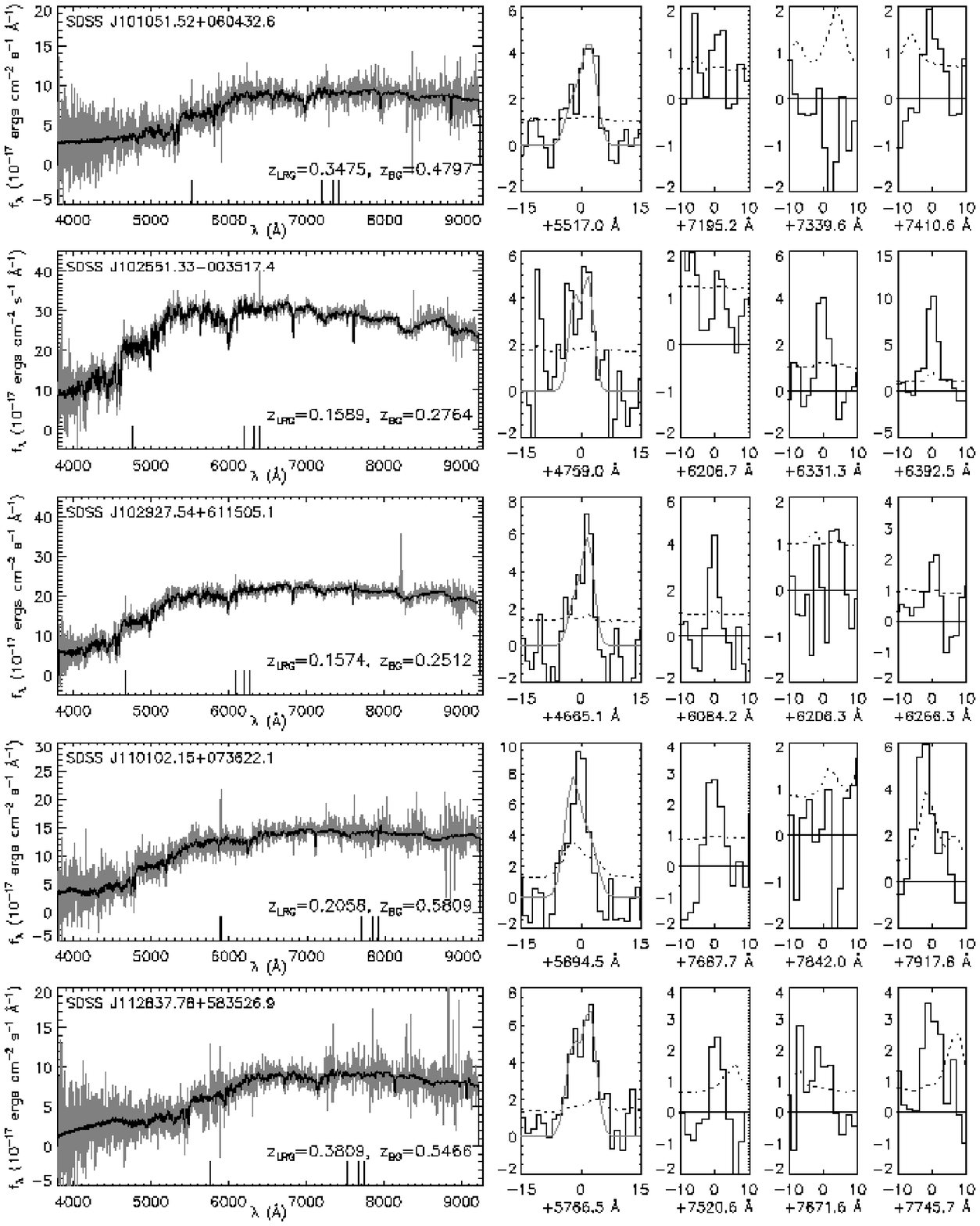}}
%\scalebox{1.15}{\plotone{bolton.fig2d.eps}}
\caption{Continued}
\end{figure*}

\addtocounter{figure}{-1}
\begin{figure*}
\scalebox{1.15}{\plotone{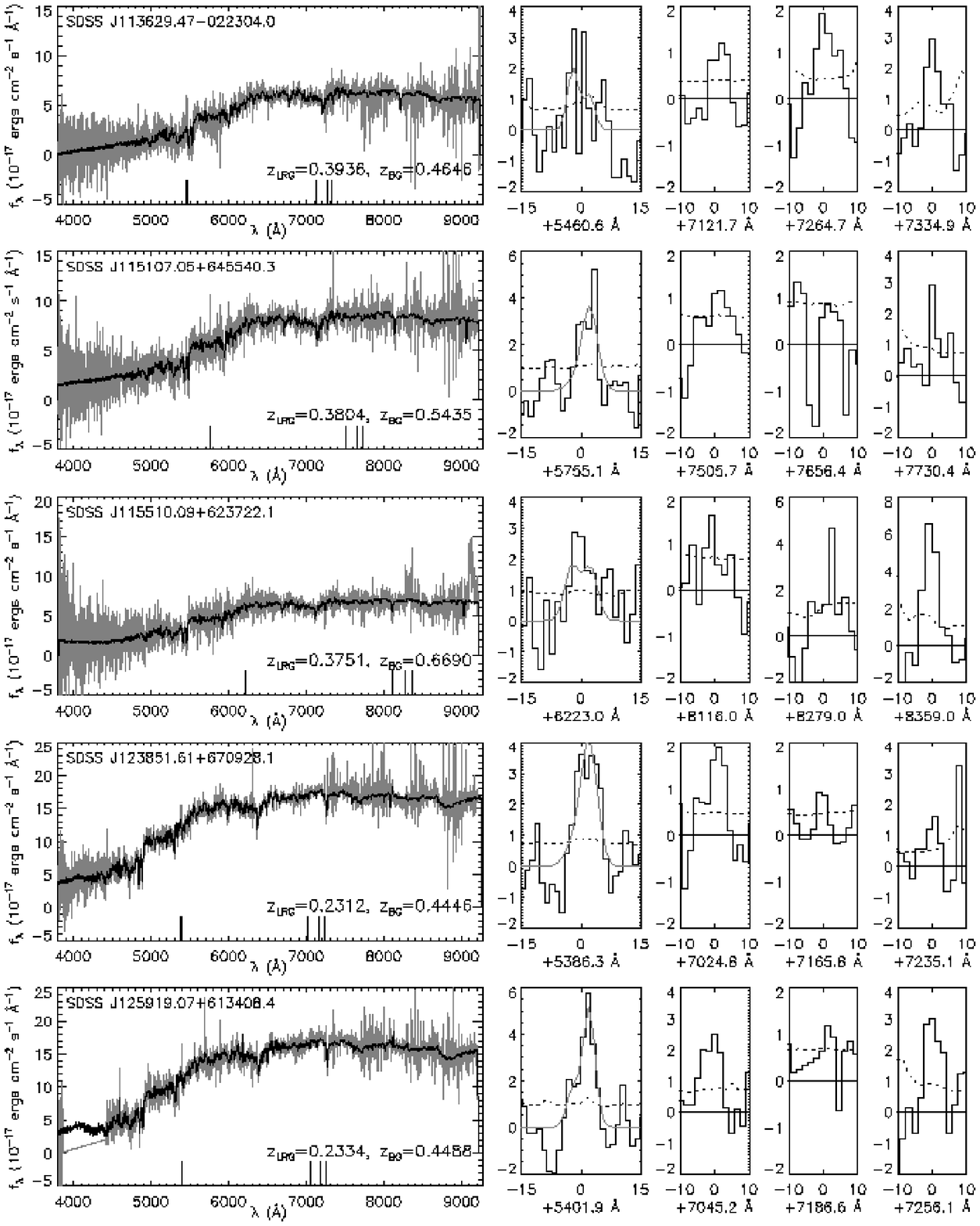}}
%\scalebox{1.15}{\plotone{bolton.fig2e.eps}}
\caption{Continued}
\end{figure*}

\addtocounter{figure}{-1}
\begin{figure*}
\scalebox{1.15}{\plotone{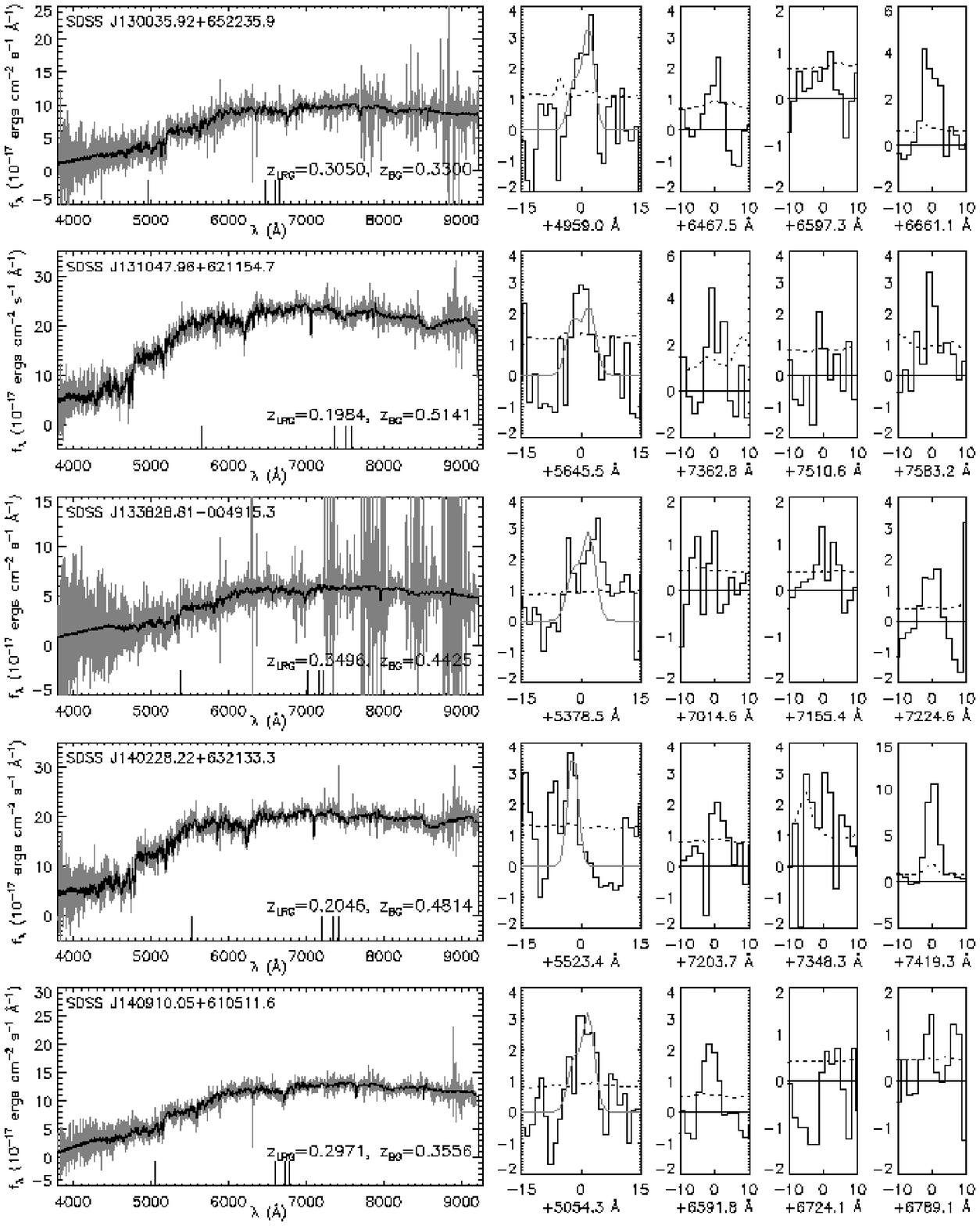}}
%\scalebox{1.15}{\plotone{bolton.fig2f.eps}}
\caption{Continued}
\end{figure*}

\addtocounter{figure}{-1}
\begin{figure*}
\scalebox{1.15}{\plotone{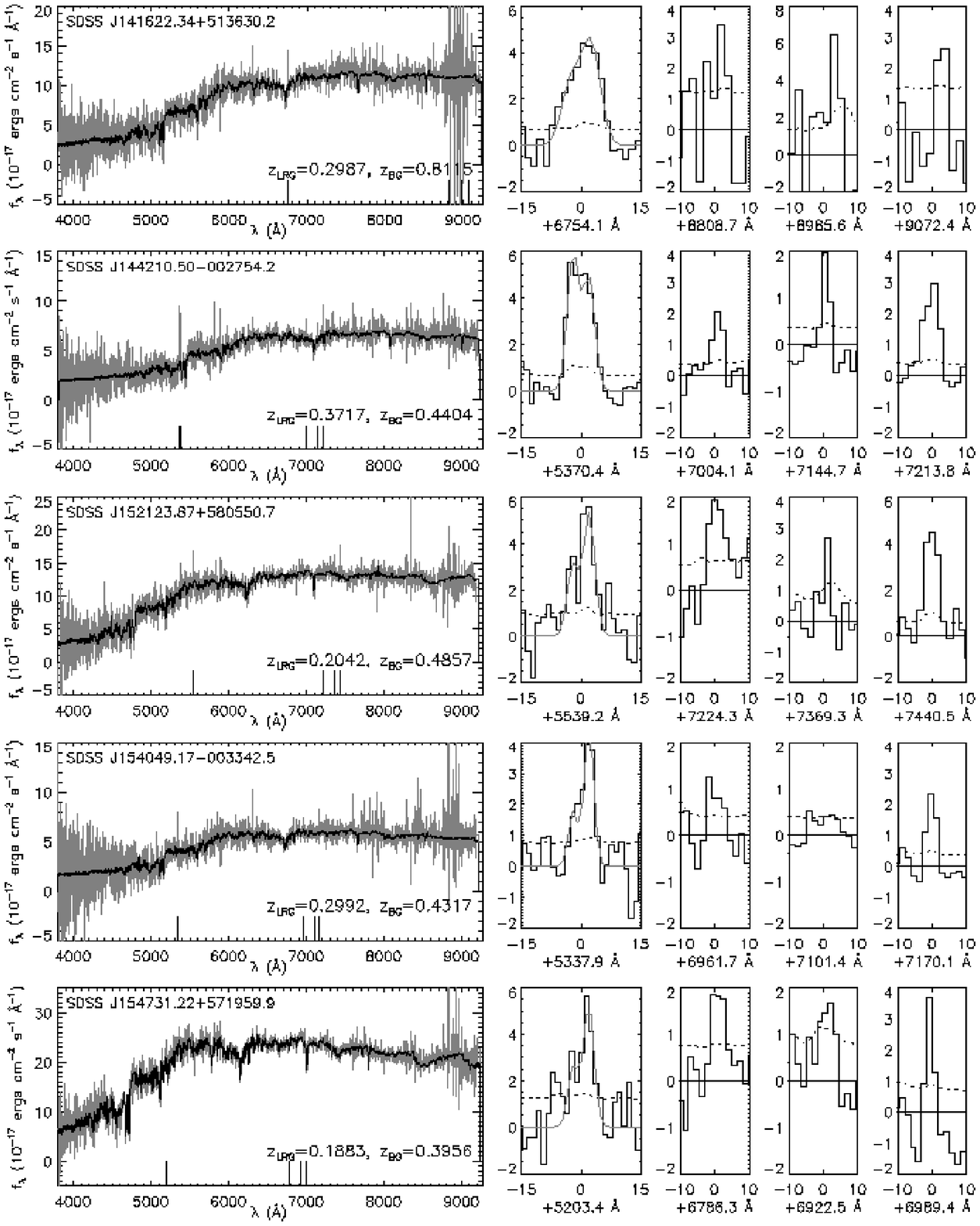}}
%\scalebox{1.15}{\plotone{bolton.fig2g.eps}}
\caption{Continued}
\end{figure*}

\addtocounter{figure}{-1}
\begin{figure*}
\scalebox{1.15}{\plotone{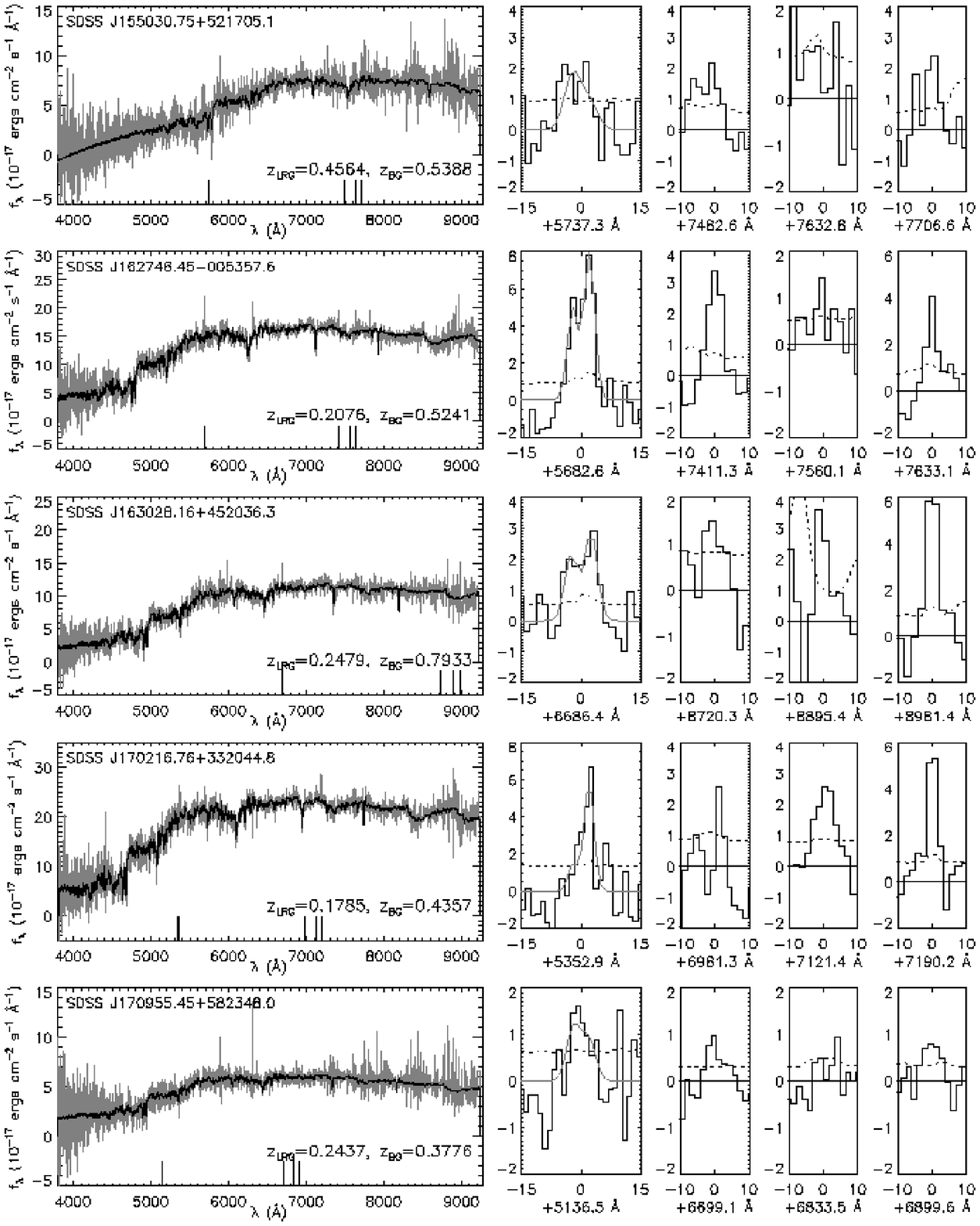}}
%\scalebox{1.15}{\plotone{bolton.fig2h.eps}}
\caption{Continued}
\end{figure*}

\addtocounter{figure}{-1}
\begin{figure*}
\scalebox{1.15}{\plotone{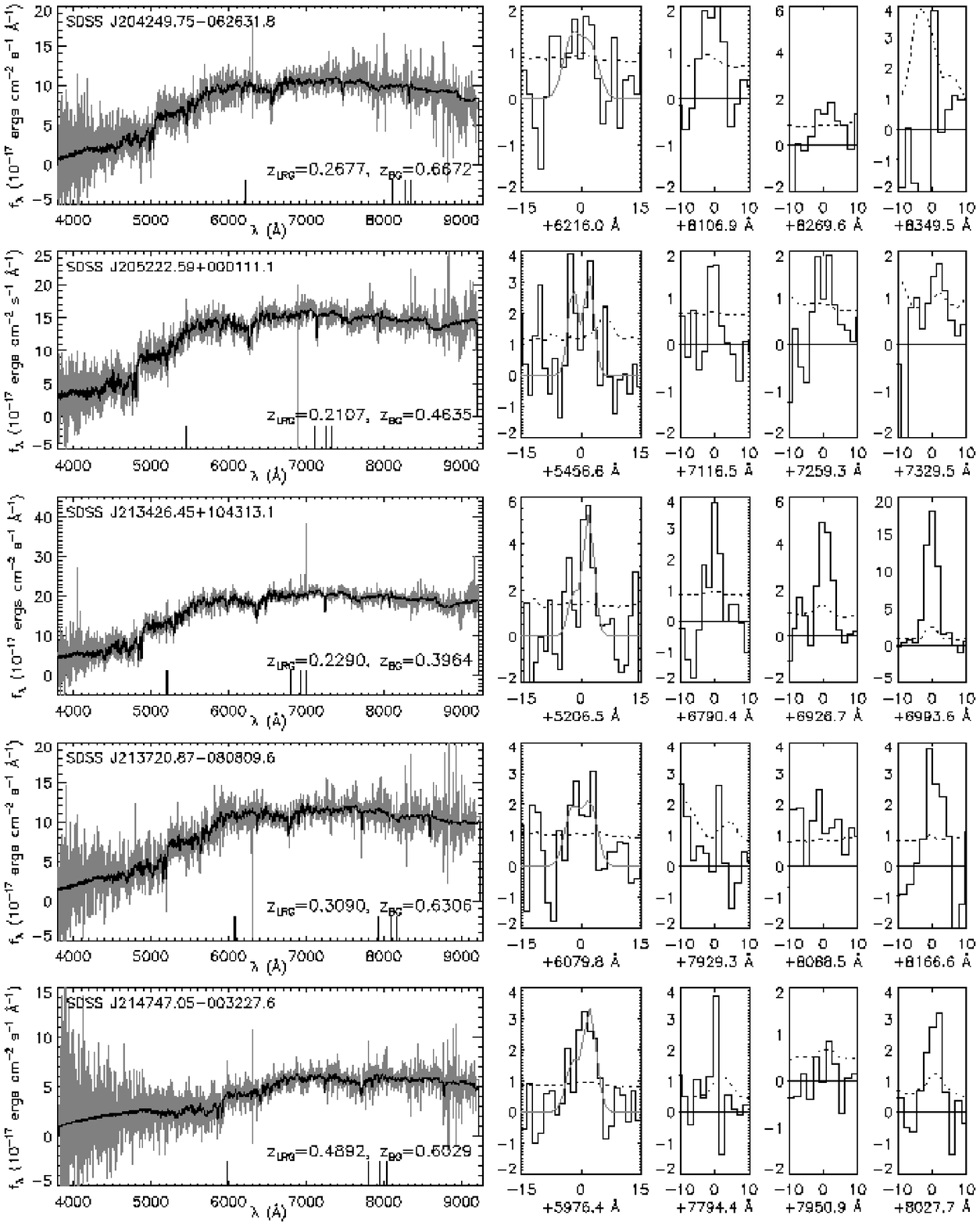}}
%\scalebox{1.15}{\plotone{bolton.fig2i.eps}}
\caption{Continued}
\end{figure*}

\addtocounter{figure}{-1}
\begin{figure*}
\scalebox{1.15}{\plotone{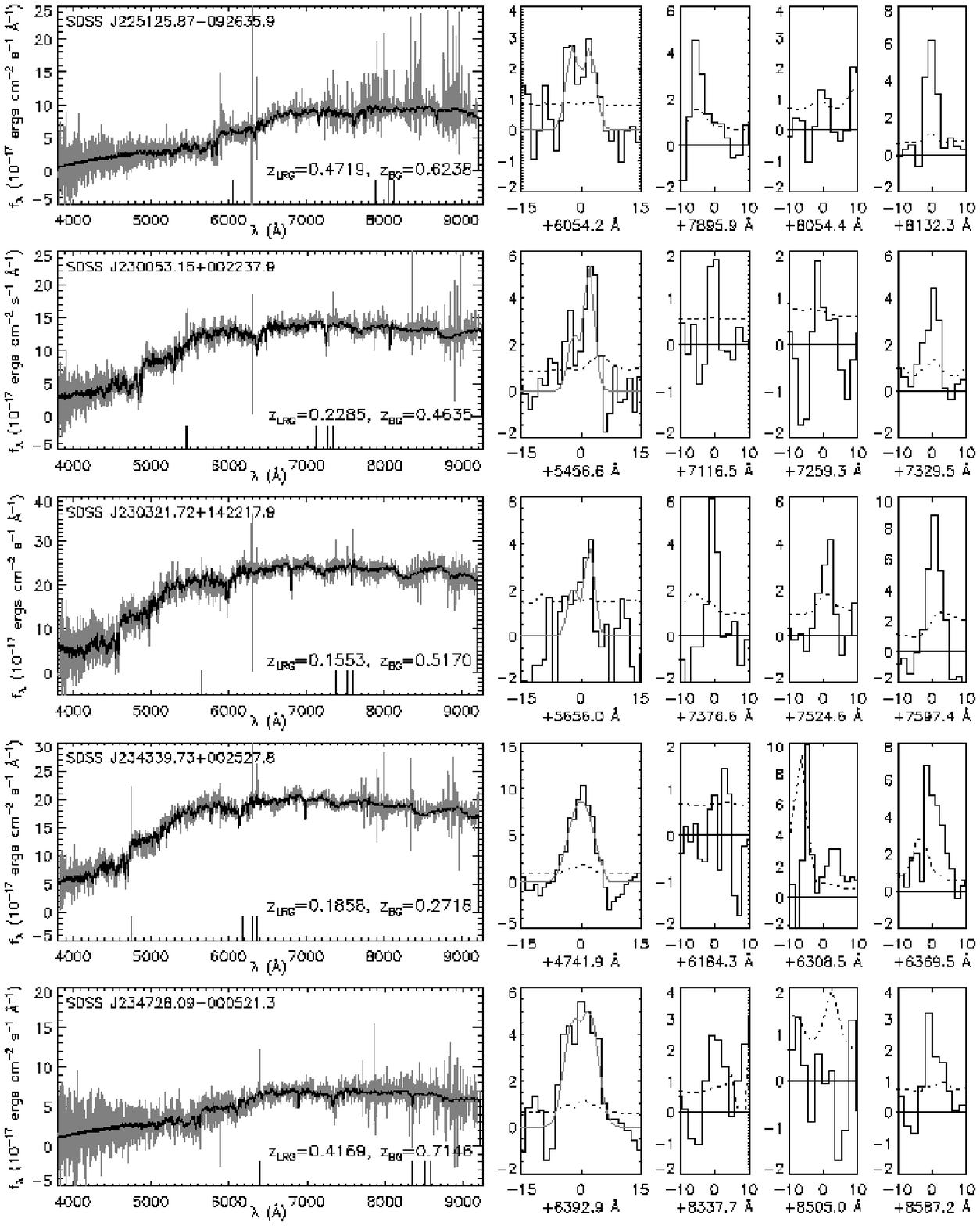}}
%\scalebox{1.15}{\plotone{bolton.fig2j.eps}}
\caption{Continued}
\end{figure*}

\begin{deluxetable*}{lcccccccclc}
\tabletypesize{\footnotesize}
%\rotate
\tablewidth{0pt}
\tabcolsep=6pt
\tablecaption{Properties of Candidate Lens Systems\label{galtab}}
\tablehead{
\colhead{} & \colhead{Plate-MJD} & 
\colhead{In} & \colhead{$r$} &
\colhead{$z$} & \colhead{$\sigma_v$} &
\colhead{$z$} & \colhead{$\Delta \theta$\tablenotemark{b}}
 & \colhead{[O~{\sc ii}] $\lambda \lambda$}
 & \colhead{Other}  & \colhead{Total} \\
\colhead{System Name\tablenotemark{a}} & 
\colhead{-Fiber} & \colhead{DR1?} & \colhead{(de Vauc.)} &
\colhead{(LRG)} & \colhead{(km s$^{-1}$)} &
\colhead{(BG)} & \colhead{($\arcsec$)}
 & \colhead{Flux\tablenotemark{c}}
 & \colhead{Lines\tablenotemark{d}}
 & \colhead{$S/N$}}
\startdata
\objectname{SDSS J002431.96$+$003123.1}\tablenotemark{e} 
& 0390-51900-589 & Y & 
18.48$\pm$0.02 & 0.3979 & 286$\pm$19\tablenotemark{f} & 
0.5952 & 1.39 & 20 $\pm$  2\phn & d,e & 11.6 \\
\objectname{SDSS J003524.12$+$002400.2}\tablenotemark{e} 
& 0689-52262-632 & N & 
17.88$\pm$0.01 & 0.2017 & 176$\pm$21\phm{\tablenotemark{f}} & 
0.2708 & 0.43 & 19 $\pm$  3\phn & c,e,g,h,i & 10.8 \\
\objectname{SDSS J003753.21$-$094220.1}\tablenotemark{e} 
& 0655-52162-392 & Y & 
16.79$\pm$0.01 & 0.1955 & 279$\pm$10\phm{\tablenotemark{f}} & 
0.6322 & 2.94 & 23 $\pm$  4\phn & a,c,d,e & 16.0 \\
\objectname{SDSS J010933.73$+$150032.4}\tablenotemark{e} 
& 0422-51811-508 & Y & 
18.26$\pm$0.01 & 0.2939 & 251$\pm$19\tablenotemark{f} & 
0.5248 & 1.47 & 19 $\pm$  3\phn & c,d,e & 10.2 \\
\objectname{SDSS J011123.49$+$000012.7} & 0694-52209-076 & N & 
19.18$\pm$0.03 & 0.4516 & \phn  86$\pm$25\tablenotemark{f} & 
0.5011 & 0.04 & 19 $\pm$  2\phn & c,e & 11.6 \\
\objectname{SDSS J021652.53$-$081345.4}\tablenotemark{e} 
& 0668-52162-428 & Y & 
17.33$\pm$0.01 & 0.3317 & 333$\pm$23\tablenotemark{f} & 
0.5235 & 2.13 & 25 $\pm$  4\phn & c,d & \phn  8.7 \\
\objectname{SDSS J032454.50$-$011029.1}\tablenotemark{e} 
& 0414-51869-244 & 
Y\tablenotemark{g} & 
18.85$\pm$0.03 & 0.4453 & 267$\pm$35\tablenotemark{f} & 
0.6240 & 1.03 & 22 $\pm$  2\phn & b,c,d,e & 16.5 \\
\objectname{SDSS J073728.44$+$321618.6} & 0541-51959-145 & N & 
17.85$\pm$0.01 & 0.3223 & 338$\pm$16\phm{\tablenotemark{f}} & 
0.5812 & 2.67 & 62 $\pm$  5\phn & a,b,c,d,e & 24.7 \\
\objectname{SDSS J080532.29$+$303712.7} & 0860-52319-452 & N & 
17.95$\pm$0.01 & 0.3200 & 334$\pm$17\phm{\tablenotemark{f}} & 
0.3809 & 0.94 & 54 $\pm$  4\phn & c,e,g,h & 17.0 \\
\objectname{SDSS J085038.70$+$020155.2} & 0468-51912-362 & Y & 
18.70$\pm$0.02 & 0.4114 & 251$\pm$41\tablenotemark{f} & 
0.5672 & 0.88 & 12 $\pm$  4\phn & d,e & \phn  6.2 \\
\objectname{SDSS J091205.31$+$002901.1} & 0472-51955-429 & Y & 
16.08$\pm$0.01 & 0.1642 & 326$\pm$12\phm{\tablenotemark{f}} & 
0.3240 & 2.90 & 26 $\pm$  4\phn & b,c,e,g & 11.5 \\
\objectname{SDSS J092857.33$+$440059.1} & 0870-52325-465 & N & 
17.76$\pm$0.01 & 0.2908 & 198$\pm$24\tablenotemark{f} & 
0.4538 & 0.75 & 29 $\pm$  4\phn & d,e & \phn  9.4 \\
\objectname{SDSS J093543.93$-$000334.8} & 0476-52314-177 & N & 
17.52$\pm$0.01 & 0.3475 & 396$\pm$35\tablenotemark{f} & 
0.4670 & 2.10 & 15 $\pm$  3\phn & c,e & \phn  8.8 \\
\objectname{SDSS J095629.78$+$510006.3} & 0902-52409-068 & N & 
16.84$\pm$0.01 & 0.2405 & 334$\pm$15\phm{\tablenotemark{f}} & 
0.4700 & 2.94 & 31 $\pm$  3\phn & c,e & 11.5 \\
\objectname{SDSS J101051.52$+$060432.6} & 0996-52641-106 & N & 
17.99$\pm$0.01 & 0.3475 & 183$\pm$22\tablenotemark{f} & 
0.4797 & 0.48 & 18 $\pm$  3\phn & c,e & \phn  7.9 \\
\objectname{SDSS J102551.32$-$003517.4} & 0272-51941-151 & Y & 
16.10$\pm$0.01 & 0.1589 & 264$\pm$11\phm{\tablenotemark{f}} & 
0.2764 & 1.64 & 18 $\pm$  3\phn & d,e,g,i & 14.6 \\
\objectname{SDSS J102927.54$+$611505.0} & 0772-52375-140 & N & 
16.12$\pm$0.01 & 0.1574 & 228$\pm$14\phm{\tablenotemark{f}} & 
0.2512 & 1.08 & 31 $\pm$  5\phn & c,e,f,g,h,i & 18.7 \\
\objectname{SDSS J110102.15$+$073622.0} & 1002-52646-504 & N & 
17.27$\pm$0.01 & 0.2058 & 217$\pm$17\tablenotemark{f} & 
0.5809 & 1.65 & 52 $\pm$ 11 & c,e & \phn  6.1 \\
\objectname{SDSS J112837.77$+$583526.8} & 0951-52398-036 & N & 
18.19$\pm$0.01 & 0.3809 & 223$\pm$28\tablenotemark{f} & 
0.5466 & 0.78 & 29 $\pm$  4\phn & c,d,e & 11.2 \\
\objectname{SDSS J113629.47$-$022303.9} & 0328-52282-350 & N & 
18.84$\pm$0.02 & 0.3936 & 321$\pm$27\tablenotemark{f} & 
0.4646 & 0.81 & \phn  9 $\pm$  3\phn & c,d,e & \phn  8.9 \\
\objectname{SDSS J115107.04$+$645540.3} & 0598-52316-477 & N & 
18.29$\pm$0.01 & 0.3804 & 260$\pm$27\tablenotemark{f} & 
0.5435 & 1.05 & 14 $\pm$  3\phn & c,e & \phn  6.5 \\
\objectname{SDSS J115510.09$+$623722.1} & 0777-52320-501 & N & 
18.04$\pm$0.02 & 0.3751 & 303$\pm$34\tablenotemark{f} & 
0.6690 & 2.08 & 16 $\pm$  4\phn & d,e & \phn  7.5 \\
\objectname{SDSS J123851.61$+$670928.1} & 0494-51915-074 & Y & 
17.10$\pm$0.01 & 0.2312 & 238$\pm$10\phm{\tablenotemark{f}} & 
0.4446 & 1.47 & 16 $\pm$  2\phn & c,e & 10.5 \\
\objectname{SDSS J125919.07$+$613408.3} & 0783-52325-279 & N & 
17.33$\pm$0.01 & 0.2334 & 253$\pm$16\phm{\tablenotemark{f}} & 
0.4488 & 1.67 & 23 $\pm$  3\phn & c,e & 10.6 \\
\objectname{SDSS J130035.91$+$652235.9} & 0602-52072-332 & Y & 
18.40$\pm$0.01 & 0.3050 & 300$\pm$24\tablenotemark{f} & 
0.3300 & 0.36 & 16 $\pm$  3\phn & c,e,g,j & 12.0 \\
\objectname{SDSS J131047.95$+$621154.7} & 0784-52327-350 & N & 
16.68$\pm$0.01 & 0.1984 & 226$\pm$12\phm{\tablenotemark{f}} & 
0.5141 & 1.71 & 16 $\pm$  5\phn & c,e & \phn  6.6 \\
\objectname{SDSS J133828.80$-$004915.2} & 0298-51955-001 & Y & 
18.02$\pm$0.02 & 0.3496 & 111$\pm$12\tablenotemark{f} & 
0.4425 & 0.14 & 18 $\pm$  3\phn & d,e & \phn  8.4 \\
\objectname{SDSS J140228.22$+$632133.3} & 0605-52353-503 & N & 
16.91$\pm$0.01 & 0.2046 & 267$\pm$17\phm{\tablenotemark{f}} & 
0.4814 & 2.23 & 10 $\pm$  3\phn & d,e & \phn  9.6 \\
\objectname{SDSS J140910.04$+$610511.5} & 0606-52365-315 & N & 
17.52$\pm$0.01 & 0.2971 & 428$\pm$21\tablenotemark{f} & 
0.3556 & 1.60 & 17 $\pm$  3\phn & c,e,g,h,j & 13.9 \\
\objectname{SDSS J141622.33$+$513630.2} & 1045-52725-464 & N & 
17.78$\pm$0.01 & 0.2987 & 240$\pm$25\tablenotemark{f} & 
0.8115 & 1.92 & 26 $\pm$  2\phn & c,d & \phn  7.7 \\
\objectname{SDSS J144210.49$-$002754.1} & 0307-51663-065 & Y & 
18.34$\pm$0.01 & 0.3717 & 349$\pm$28\tablenotemark{f} & 
0.4404 & 0.99 & 40 $\pm$  3\phn & c,d,e & 15.8 \\
\objectname{SDSS J152123.87$+$580550.6} & 0615-52347-311 & 
Y\tablenotemark{g} & 
17.25$\pm$0.01 & 0.2042 & 174$\pm$16\phm{\tablenotemark{f}} & 
0.4857 & 0.96 & 30 $\pm$  4\phn & c,e & 12.1 \\
\objectname{SDSS J154049.17$-$003342.5} & 0315-51663-143 & Y & 
17.63$\pm$0.01 & 0.2992 & 300$\pm$35\tablenotemark{f} & 
0.4317 & 1.46 & 19 $\pm$  3\phn & b,c,e & \phn  9.9 \\
\objectname{SDSS J154731.22$+$571959.8} & 0617-52072-561 & Y & 
16.82$\pm$0.01 & 0.1883 & 254$\pm$12\phm{\tablenotemark{f}} & 
0.3956 & 1.85 & 27 $\pm$  4\phn & c,e,g,h & \phn  8.5 \\
\objectname{SDSS J155030.75$+$521705.0} & 0618-52049-123 & Y & 
18.38$\pm$0.02 & 0.4564 & 345$\pm$52\tablenotemark{f} & 
0.5388 & 0.92 & \phn  8 $\pm$  2\phn & c,e & \phn  6.8 \\
\objectname{SDSS J162746.44$-$005357.5} & 0364-52000-084 & Y & 
17.50$\pm$0.01 & 0.2076 & 290$\pm$14\phm{\tablenotemark{f}} & 
0.5241 & 2.76 & 29 $\pm$  3\phn & c,e & 12.5 \\
\objectname{SDSS J163028.16$+$452036.2} & 0626-52057-518 & Y & 
17.27$\pm$0.01 & 0.2479 & 276$\pm$16\phm{\tablenotemark{f}} & 
0.7933 & 2.81 & 21 $\pm$  3\phn & d,e & 10.1 \\
\objectname{SDSS J170216.76$+$332044.7} & 0973-52426-464 & N & 
16.85$\pm$0.01 & 0.1785 & 256$\pm$14\phm{\tablenotemark{f}} & 
0.4357 & 2.12 & 23 $\pm$  5\phn & d,e & \phn  9.2 \\
\objectname{SDSS J170955.44$+$582348.0} & 0353-51703-121 & Y & 
17.39$\pm$0.01 & 0.2437 & \phn  78$\pm$20\tablenotemark{f} & 
0.3776 & 0.12 & \phn  9 $\pm$  2\phn & c,e,g & \phn  8.5 \\
\objectname{SDSS J204249.75$-$062631.8} & 0635-52145-290 & Y & 
18.04$\pm$0.01 & 0.2677 & 264$\pm$28\tablenotemark{f} & 
0.6672 & 2.23 & 10 $\pm$  3\phn & c,d & \phn  6.0 \\
\objectname{SDSS J205222.58$+$000111.1}\tablenotemark{e} 
& 0982-52466-602 & N & 
17.53$\pm$0.01 & 0.2107 & 320$\pm$17\phm{\tablenotemark{f}} & 
0.4635 & 3.04 & 10 $\pm$  2\phn & c,d & \phn  6.5 \\
\objectname{SDSS J213426.45$+$104313.1} & 0731-52460-165 & N & 
16.89$\pm$0.01 & 0.2290 & 224$\pm$14\phm{\tablenotemark{f}} & 
0.3964 & 1.14 & 14 $\pm$  3\phn & c,d,e,g & 15.3 \\
\objectname{SDSS J213720.87$-$080809.5}\tablenotemark{e} 
& 0641-52199-021 & Y & 
17.87$\pm$0.01 & 0.3090 & 201$\pm$21\tablenotemark{f} & 
0.6306 & 1.08 & 11 $\pm$  2\phn & d,e & \phn  7.9 \\
\objectname{SDSS J214747.04$-$003227.6}\tablenotemark{e} 
& 0371-52078-307 & N & 
18.66$\pm$0.02 & 0.4892 & 379$\pm$36\tablenotemark{f} & 
0.6029 & 1.35 & 12 $\pm$  2\phn & a,c,e & \phn  8.3 \\
\objectname{SDSS J225125.87$-$092635.8}\tablenotemark{e} 
& 0724-52254-277 & N & 
18.33$\pm$0.02 & 0.4719 & 414$\pm$47\tablenotemark{f} & 
0.6238 & 2.09 & 14 $\pm$  2\phn & c,e & \phn  9.7 \\
\objectname{SDSS J230053.14$+$002237.9}\tablenotemark{e} 
& 0677-52606-520 & N & 
17.59$\pm$0.01 & 0.2285 & 279$\pm$17\phm{\tablenotemark{f}} & 
0.4635 & 2.13 & 19 $\pm$  3\phn & c,e & \phn  9.5 \\
\objectname{SDSS J230321.72$+$142217.9}\tablenotemark{e} 
& 0743-52262-304 & N & 
16.69$\pm$0.01 & 0.1553 & 255$\pm$16\phm{\tablenotemark{f}} & 
0.5170 & 2.51 & 13 $\pm$  3\phn & c,e & \phn  8.1 \\
\objectname{SDSS J234339.73$+$002527.7}\tablenotemark{e} 
& 0385-51877-596 & Y & 
16.54$\pm$0.01 & 0.1858 & 229$\pm$14\phm{\tablenotemark{f}} & 
0.2718 & 0.91 & 53 $\pm$  4\phn & d,e,g,h,i,j & 16.6 \\
\objectname{SDSS J234728.08$-$000521.2}\tablenotemark{e} 
& 0684-52523-311 & N & 
18.42$\pm$0.02 & 0.4169 & 404$\pm$59\tablenotemark{f} & 
0.7146 & 3.47 & 30 $\pm$  3\phn & c,e & 12.7 \\
\tablenotetext{a}{~Names give truncated J2000 RA and Dec
in the format HHMMSS.ss$\pm$DDMMSS.s.}
\tablenotetext{b}{~$\Delta \theta$ column gives an estimated angular
lensing scale for each system, calculated as described in the text.}
\tablenotetext{c}{~\oii\ line flux in units of
10$^{-17}$ ergs cm$^{-2}$ s$^{-1}$.}
\tablenotetext{d}{~Other lines detected at 2.5-$\sigma$ or greater.
Letters specify lines as follows: a~=~\hd, b~=~\hg, c~=~\hb,
d~=~\oiiia, e~=~\oiiib, f~=~\niia, g~=~\ha, h~=~\niib,
i~=~\siia, j~=~\siib.}
\tablenotetext{e}{~See \S~\ref{lensornot} for remarks on
follow-up imaging of these systems.}
\tablenotetext{f}{~Velocity dispersion ($\sigma_v$) value
possibly unreliable: either too high ($>$ 420 km s$^{-1}$),
too low to be properly resolved ($<$ 100 km s$^{-1}$),
or LRG spectrum
median signal-to-noise per pixel too low ($<$ 10).}
\tablenotetext{g}{~SDSS-DR1 contains lower signal-to-noise
spectra of these systems from earlier observations, with
Plate-MJD-Fiber values of 414-51901-266 (for SDSS J0324)
and 612-52079-607 (for SDSS J1521).}
\enddata
\label{galtable}
\end{deluxetable*}

In this section we present our
catalog of 49 candidate lensed star-forming
galaxies selected to have
\oii\ emission at $S/N$ of 3 or higher and emission
from two out of the three of \hb, \oiiia, and \oiiib\ each
at $S/N$ of 2.5 or higher,
at a redshift significantly greater than that of the
primary target LRG.  Table~\ref{galtable}
lists various properties of the candidate lens systems.
LRG $r$ de Vaucouleurs model magnitudes are
determined from SDSS imaging and photometric reduction.
LRG redshifts and velocity dispersions are as
provided by {\tt specBS}; the software
fits for velocity dispersions
$\sigma_v$ using a set of 24 stellar eigenspectra
derived from a PCA of the ELODIE
spectral library \citep{elodie}.  We report
all $\sigma_v$ values from the database, although
some are likely unreliable; see the notes
of Table~\ref{galtable}.  We also report
emission-line redshifts
of the detected background galaxies.
Using the observed
LRG and background redshifts and the observed
LRG $\sigma_v$, and assuming a
singular isothermal sphere (SIS) LRG
luminous$+$dark matter distribution,
we calculate a ``best guess''
for the angular scale of any lensing that might
be present in these systems as
$\Delta \theta = 8 \pi (\sigma_v^2 / c^2) (D_{LS} /D_S)$.
($D_{LS}$ and $D_S$ are
angular-diameter distances from lens to source
and from observer to source\footnote{
Throughout this paper, we assume a
cosmology of
$(\Omega_M, \Omega_{\Lambda}) = (0.3, 0.7)$
and a Hubble constant $H_0$ of
70 km s$^{-1}$ Mpc$^{-1}$.
Since $\Delta \theta$ is proportional
to a distance ratio, it is independent
of $H_0$.}).
This is the separation
between the two images of a strongly lensed object
in the SIS model; it is also the radius of
the strong-lensing region of the image plane, and
twice the radius of ring images of compact sources
directly behind the lens \citep[for example]{nb96}.
Finally, for each candidate system
we report the detected background
\oii\ line flux from the best-fit double Gaussian
profile, a list of other lines detected at $S/N > 2.5$,
and the total $S/N$ of the background galaxy emission-line
fit as defined in \S~\ref{select}.
The reported background line fluxes should be viewed
with suspicion: the spatial alignments of the
background galaxies are unknown,
and the spectroscopic fibers will not necessarily
have captured all of their light.
In addition, most of the reduced candidate LRG spectra
have been rescaled based on smear exposures that
attempt to account for LRG flux falling outside
the $3\arcsec$ fiber \citep{edr};
we have removed this correction
from the tabulated line fluxes, as it has no physical
relation to the background emission-line galaxies.
Figure~\ref{spectra} shows the SDSS discovery spectra
and best-fit model spectra, along with close-up
views of the residual (data $-$ model) spectra in
the wavelength ranges corresponding
to redshifted background \oii, \hb, \oiiia, and \oiiib.
Smear corrections have not been removed from these plots.

\begin{figure}
\centerline{\scalebox{1.3}{\plotone{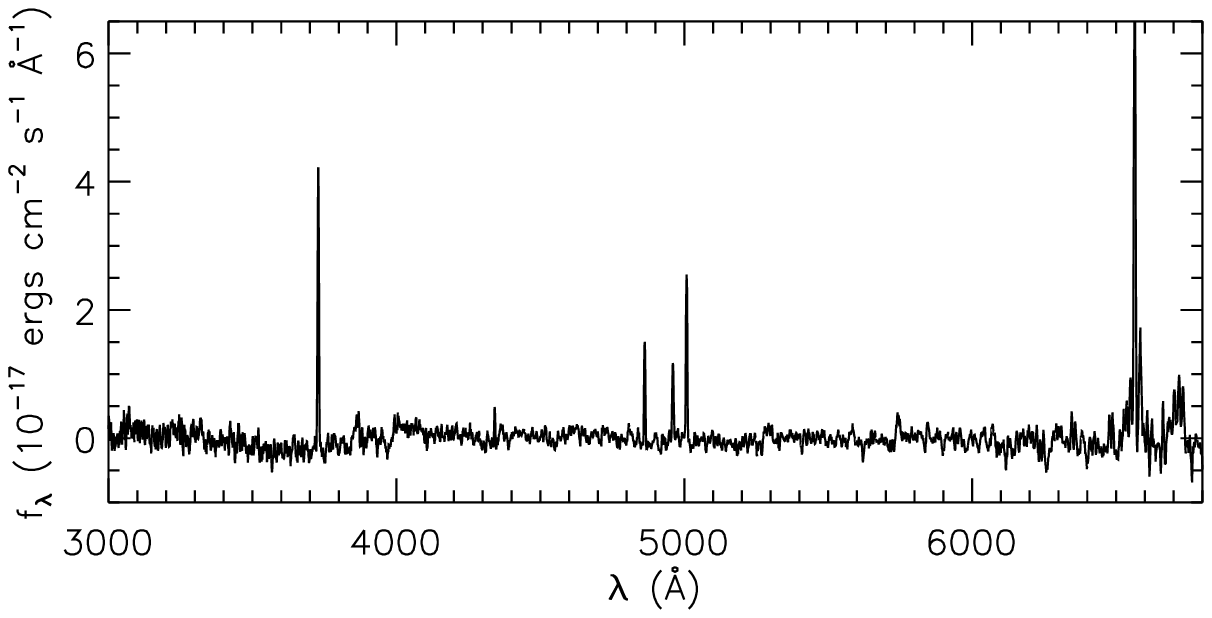}}}
\caption{Median rest-frame spectrum of background galaxies
detected in our residual LRG spectra, created as described
in the text and smoothed with a 5-pixel boxcar.  Note the
continuum break at 4000\AA\ and the absorption
feature at 5270\AA, which are not
discernible in the individual residual spectra.
Also evident are the emission lines \hg, \ha,
\niia, \niib, \siia, and \siib, in addition to the
emission lines for which we select.
\label{medspec}}
\end{figure}

Although we detect line emission clearly,
evidence of background galaxy {\em continuum} in the
residual spectra of our candidate systems is
scarce.  This is not surprising, for three reasons.
One, the LRG sample was selected for particular
broadband color and luminosity, and significant
background continuum would likely perturb an
LRG out of the sample.  Two, any faint background
continuum present in an LRG spectrum will largely
project onto the LRG-redshift eigenspectrum set
and low-order polynomial fit used by
{\tt specBS}, and will be subtracted
along with the LRG model when forming the residual
spectrum.  Three, these background galaxies are likely
to be high-equivalent-width star-forming systems,
and since their line fluxes are detected just above
the noise threshold, the associated continuum will
typically be lost in the noise.
Nevertheless, we may obtain a
higher signal-to-noise picture of the
background galaxies that we detect by constructing a
median spectrum as follows.
First we transform the residual spectra of our candidate
systems into units of
erg~cm$^{-2}$~s$^{-1}$~pixel$^{-1}$,
which is a redshift-independent quantity
since the SDSS pixels are of constant velocity width.
We then shift these residual spectra into the
rest frame of the background galaxy, rounded
to the nearest whole pixel, and
transform back to erg~cm$^{-2}$~s$^{-1}$~\AA$^{-1}$.
Next we renormalize
the spectra by dividing each one by
its best-fit \oii -flux value.  We then take the
median value at each pixel, and restore physical
normalization by multiplying this median spectrum
by the sample-median
best-fit \oii -flux value.
The resulting median spectrum is shown
in Figure~\ref{medspec}.  Although there is
no discernible continuum in the individual
residual spectra, we can see a 4000-\AA\
continuum break in the median
spectrum; we also see H$\gamma$, H$\alpha$,
[N~{\sc ii}] and [S~{\sc ii}] emission lines in addition
to the lines for which we select.
This gives further evidence that
we have successfully detected and
identified background emission features.

\begin{figure*}
\scalebox{1.15}{\plotone{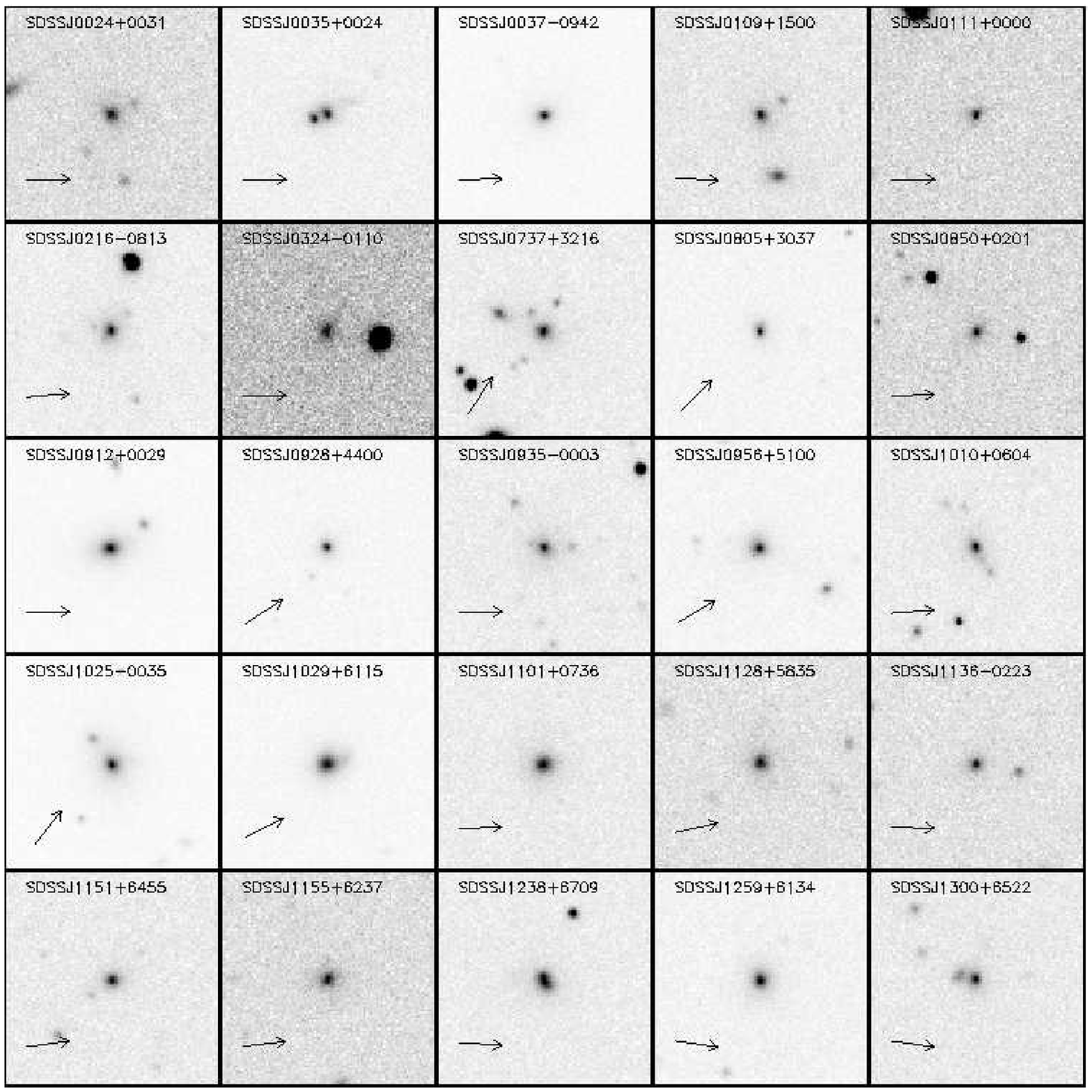}}
%\scalebox{1.15}{\plotone{bolton.fig4a.eps}}
\caption{$40\arcsec \times 40\arcsec$ SDSS $r$
postage-stamp images of LRG systems with confirmed
background emission. Gray-scaling is linear from $-3\sigma$
sky noise (white) to LRG peak brightness (black).
Arrows point North; East is $90^{\circ}$
counterclockwise from North (i.e., as seen on the sky).
\label{ims}}
\end{figure*}

\addtocounter{figure}{-1}
\begin{figure*}
\scalebox{1.15}{\plotone{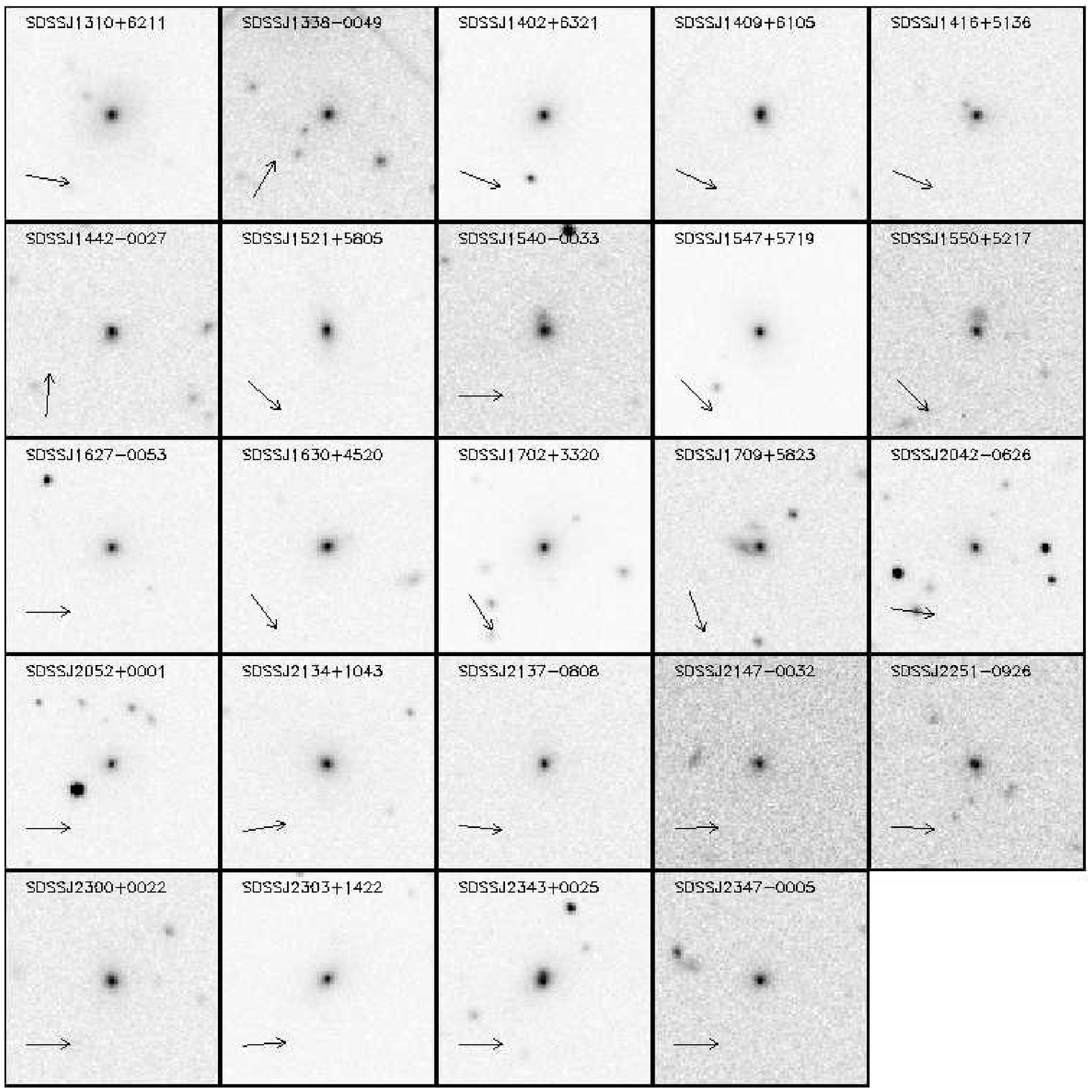}}
%\scalebox{1.15}{\plotone{bolton.fig4b.eps}}
\caption{Continued}
\end{figure*}

\subsection{Lenses or Not?}
\label{lensornot}

We have with certainty detected emission from galaxies
behind foreground LRGs.
For a system to be a strong gravitational lens, the
background galaxy must be located at sufficiently small
impact parameter relative to the LRG center.
The true incidence of lensing within our sample
can best be determined and studied with
either narrow-band imaging
or integral-field spectroscopy.
Such observations could spectrally isolate
the background line flux and resolve
it spatially to reveal any lensing morphology.
SDSS broadband imaging offers some hope
for answering the lensing question, but
in general Sloan images do not detect late-type
galaxies at redshift $z \sim 0.5$ at very
high $S/N$.
Figure~\ref{ims} shows
$40\arcsec \times 40\arcsec$ SDSS $r$
postage-stamp images centered on the candidate systems,
with linear gray-scaling from $-3$-$\sigma$ sky
noise to peak LRG surface brightness.
Evidence of significantly offset
neighboring broadband emission
is seen in some images, but it would be
difficult to rule out many systems as definite
non-lenses based on SDSS-quality images.
In the spirit of a purely spectroscopic survey,
we present as candidates
all systems selected spectroscopically.
Furthermore, we note that many of these LRGs live in
high-density group/cluster environments,
and neighboring images may be at
the LRG redshift and not the source of
the background emission that we detect.

\begin{figure*}
\plotone{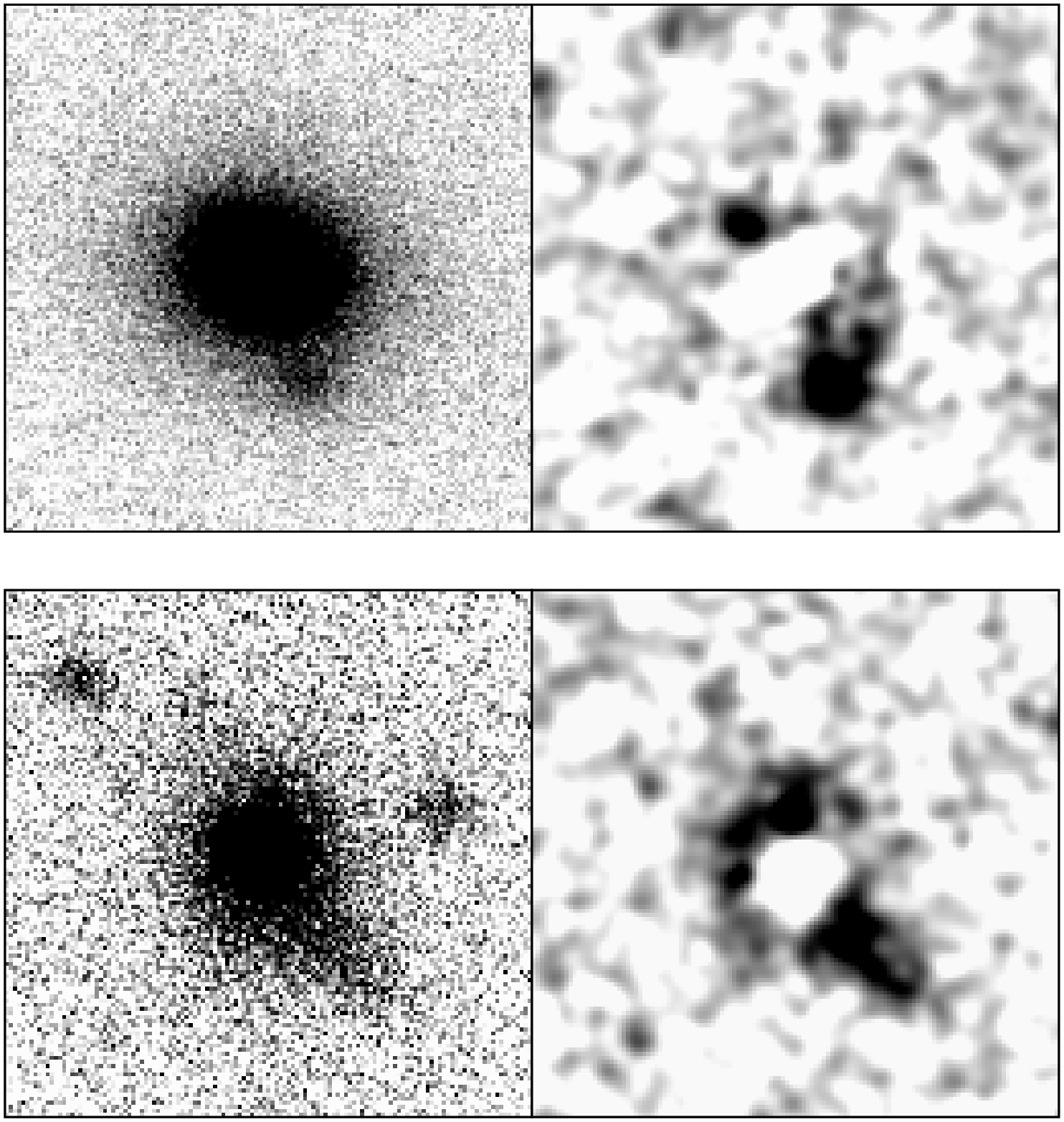}
\caption{MagIC $8 \farcs 8 \times 8 \farcs 8$
images and difference images of
lens candidates SDSSJ0037 (top) and SDSSJ0216 (bottom).
Left-hand panels show 120-s $g$-band images of the
two systems; right-hand panels show
smoothed $g - i$ difference images
created as described in the text.  Note the evidence for
multiply imaged, relatively blue objects in the difference
images.  (In these images, North is left and East is
up---reversed parity from what is seen on the sky.)
\label{diffims}}
\end{figure*}

In preparation for integral-field
spectroscopic follow-up, we have
obtained broadband reconnaissance images
of 14 of our candidate systems
(as indicated in Table~\ref{galtable})
using the Magellan consortium's 6.5-m Clay telescope
at Las Campanas Observatory.
These short $g$- and $i$-band exposures (120s each)
were taken on the nights of 2003 August 1 and 2
with the Magellan Instant Camera (MagIC)
CCD imager and active telescope
optics \citep{sch03}.  Conditions were similar on both
nights: non-photometric due to cirrus
clouds, and with $\sim 0 \farcs 8$
median FWHM seeing.  For the majority of these systems,
the images present no significant evidence
for or against the presence of strong lensing;
we describe the exceptions here.
One system, SDSSJ0035, is almost certainly not
a strong lens: a relatively $g$-bright galaxy
can be seen clearly in the images, approximately
$2 \farcs 6$ offset from the LRG.  This
far exceeds the anticipated lensing scale
of about $0 \farcs 4$ reported in
Table~\ref{galtable}.  The images of
SDSSJ2147 reveal a bluer galaxy approximately
$1 \farcs 3$ to the South, about at the
limit of the estimated lensing scale.
We see no particular evidence for
strong lensing, but deeper observation
of this system could be of interest.
Two systems, SDSSJ0037 and
SDSSJ0216, show evidence of
multiple imaging.  Figure~\ref{diffims}
shows $g$-band images of these two
systems, along with difference images
created by smoothing aligned $i$-band images with a
Gaussian kernel in order to suppress noise
and match the slightly worse $g$-band seeing, then
fitting this smoothed $i$-band LRG image
to the $g$-band image and subtracting it.
The difference images are then smoothed with
a Gaussian kernel of FWHM $\approx 0 \farcs 4$
in order to bring out coherent features;
they show strong evidence of
multiply imaged, relatively blue sources.
SDSSJ0037 seems likely to be a double-image lens.
It is tempting to identify quadruple-image
morphology in the SDSSJ0216 difference image,
but the level of noise recommends caution.
Narrow-band imaging or integral-field spectroscopy
should permit quantitative study of these systems
and of those for which broadband imaging
is inconclusive.  {\sl HST} imaging of any
confirmed lenses using the
narrow-band ramp filter set
of the {\sl Advanced Camera for Surveys}
could also be pursued
to obtain a highly resolved picture of
lensed line-emitting regions and permit
even more detailed study of the lensing mass
distributions.

The degree of lensing that will be present in
a particular system will depend on the impact parameter
(angular offset)
of the background galaxy in the source plane
relative to the center of the LRG:
the smaller the impact parameter, the more lensing
will be seen.
To obtain a rough estimate of the number of lenses within
our sample, we can guess at the unlensed
surface brightness distribution of our background
galaxies, compute lensed images under an assumed lens
model, smear to account for seeing, and integrate
over the $3\arcsec$-diameter SDSS spectroscopic
fiber, then compare to the [O~{\sc ii}] line
flux values that we have observed.
To interpret the results in terms
of lensing probabilities we must also invoke an \oii\
luminosity function (LF).  Appendix~\ref{lensprob}
describes the details of a lensing probability
calculation of this nature that
we have carried out.
The results suggest that a total of 19 out
of 49 systems are likely to be strong
lenses---that is, we expect approximately
19 systems to have source-galaxy impact parameters
less than the critical value for multiple imaging.
It is important to
recognize, though, that the ``lenses-or-not'' question
does not have as straightforward an
answer for extended sources as it does for
point sources.
Different regions of our
background galaxies will be lensed by different amounts,
and in general some but not all of the galaxy can
be multiply imaged.
It is more appropriate to
ask {\em how much} lensing is present
in any given system.

There is a sense in which the unknown degree
of lensing in our systems may prove to be of scientific
interest.  \citet{bin03} raises the concern that current
strong-lensing evidence for massive dark-matter halos
in elliptical galaxies may reflect a selection bias.
We quote directly: ``even if only a minority of
ellipticals have massive dark halos, nearly all the
observed lenses will belong to that minority.''
Our LRG sample, in contrast, has not been
selected for lensing, but rather for back-lighting.
The true incidence of lensing within our sample will
thus constitute a test of whether or not the dark halos
in currently known lenses are in fact generic
to early-type galaxies.

\begin{figure}
\centerline{\scalebox{1.25}{\plotone{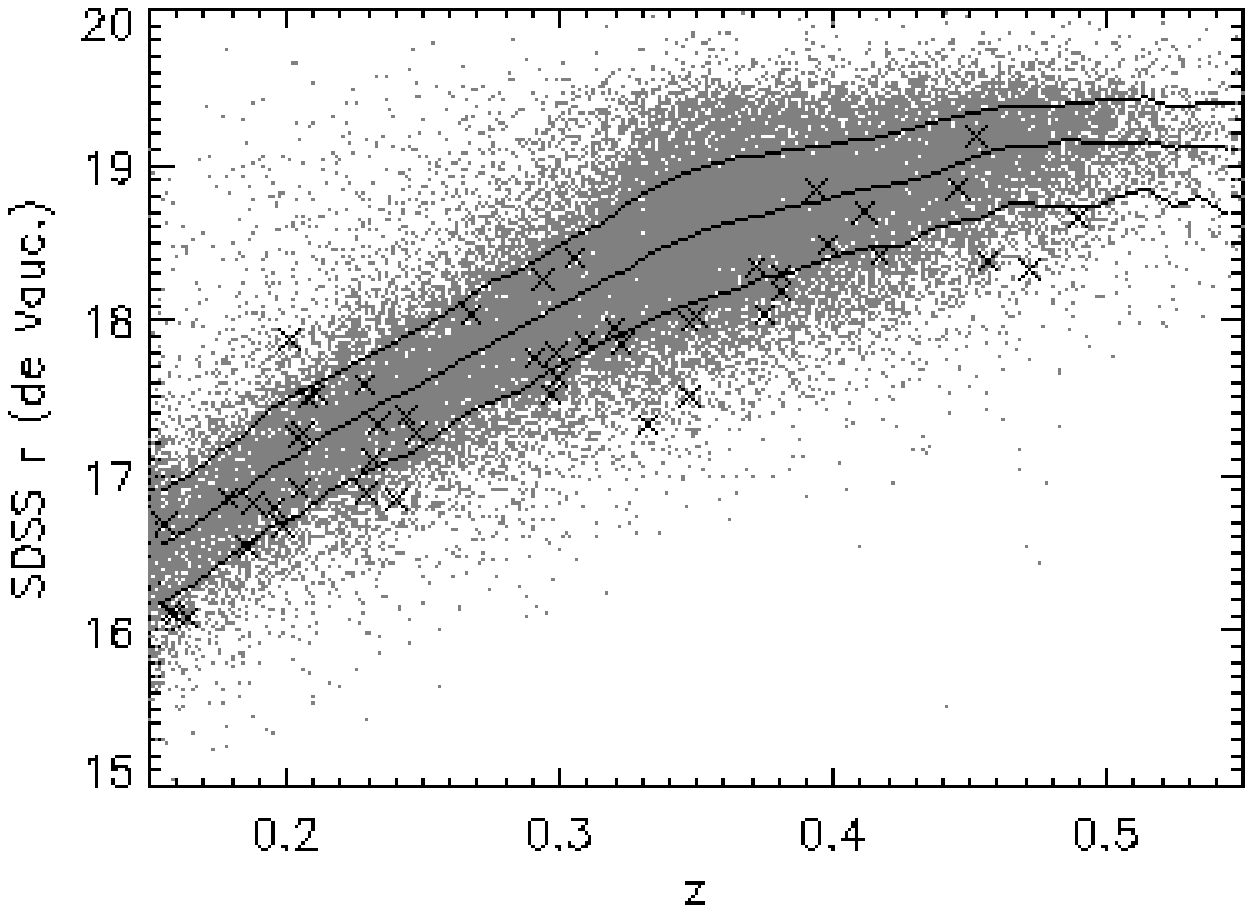}}}
%\centerline{\scalebox{1.25}{\plotone{bolton.fig6.eps}}}
\caption{Comparison of lens-candidate LRGs (crosses)
to full LRG sample (gray dots)
in the magnitude-redshift plane.  The three black lines
show the redshift-dependent median
de Vaucouleurs $r$ magnitude
of the full sample and the 84th- and 16th-percentile
$r$ values.
\label{lrgdist}}
\end{figure}

We present a comparison between our lens candidate
LRGs and the full 51,000 LRG spectroscopic sample
in Figure~\ref{lrgdist}.  The lens candidate LRGs seem
somewhat skewed towards brighter magnitudes.
The conservative interpretation is that
the broadband selection
properties of fainter LRGs are more
easily perturbed by background galaxies,
but it may also reflect a lensing signal, with
more massive galaxies providing more magnification.
If the logarithmic slope of the
underlying [O~{\sc ii}] LF is
steeper than $-1$ at the luminosities probed by
our survey, then magnified lines of sight
should show a statistically enhanced
number of [O~{\sc ii}] emitters \citep{tog84}.
\citet{hogg98} find the logarithmic [O~{\sc ii}] LF
slope to be steeper than $-1$ for line
luminosities $\ga 10^{42}$ ergs s$^{-1}$,
whereas the median observed [O~{\sc ii}] luminosity
in our sample is of order
$10^{41}$ ergs s$^{-1}$
(again, we have taken $H_0 = 70$
km s$^{-1}$ Mpc$^{-1}$).  These numbers suggest
that magnification bias is not the explanation
for the observed brightness of our lens
candidates relative to the full LRG sample, but
a more definite statement must await
follow-up observations that capture the total
background line flux.

\begin{figure}
\centerline{\scalebox{1.15}{\plotone{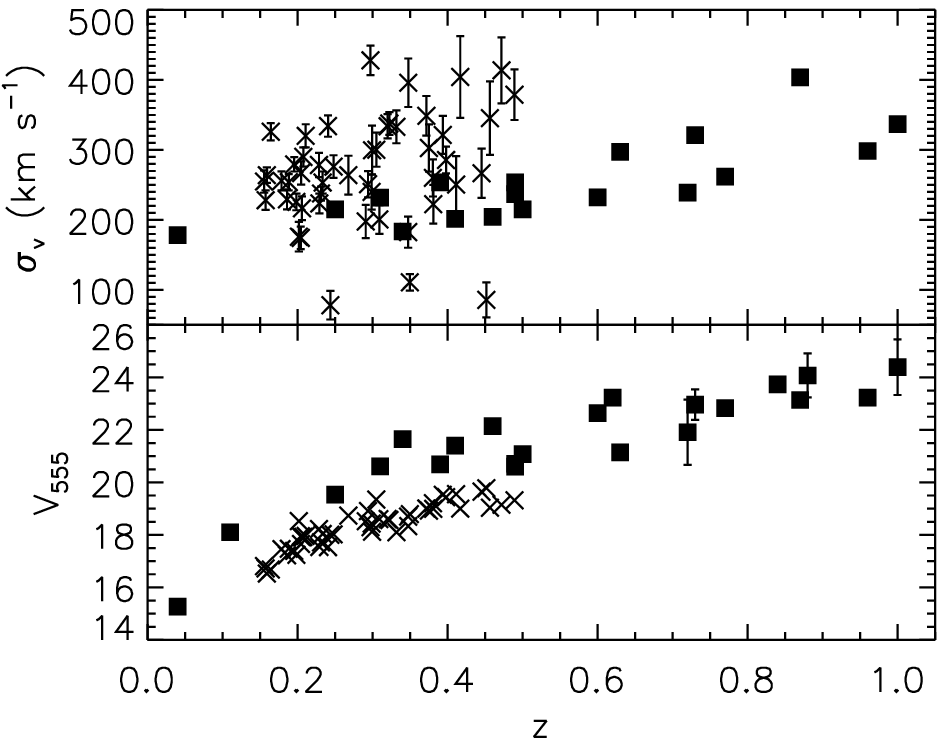}}}
\caption{Comparison of candidate LRG lens
galaxies (crosses) with known early-type lenses with
known lens redshifts (filled squares).
{\sl Top panel}: comparison of velocity dispersions $\sigma_v$
measured for LRG lens candidate galaxies and inferred from
image separations $\Delta \theta$
in known early-type lens systems with
known lens and source redshifts.
Image separations are taken from \citet{lensevol},
and conversion to velocity dispersion estimates is
made via Equation 5 of \citet{lensfund}.
For visual clarity, we omit error bars on the latter
data points; \citet{lensfund} estimate 10\% errors
in the $\Delta \theta \rightarrow \sigma_v$
conversion.
{\sl Bottom panel}: $V$-band magnitude comparison.
SDSS $g$- and $r$-band magnitudes
are converted to Johnson-Morgan $V$ via
the observed transformation of \citet{sdss2jc};
we have also corrected for
an 0.2-magnitude systematic error in the SDSS
magnitudes \citep{dr1}.
Error bars on SDSS photometric quantities are smaller
than plotted symbol sizes.
Known-lens photometric data taken from \citet{lensevol}.
All magnitudes have been corrected for galactic
extinction using \citet{sfd98} dust maps.  
\label{colormagvdispz}}
\end{figure}

Most known gravitational lens systems have been selected on the
basis of some combination of 
source properties and lens cross section,
whereas the SDSS LRG sample is selected based on
colors and magnitude.  It is therefore of interest to compare
known early-type lens galaxies to our candidate lenses,
although this is difficult since we do not currently
have truly comparable observations of the two samples.
Figure~\ref{colormagvdispz} presents
our best attempt at such a comparison
for lens velocity dispersions and apparent
magnitudes as a function of
redshift\footnote{
We plot these quantities as a function of
redshift to avoid the issue of evolutionary
and $k$ corrections.}, with
{\sl HST} known-lens data taken
from \citet{lensevol}.  The known-lens
velocity dispersions are estimated from lensed image
separations in the manner described by \citet{lensfund};
they may be systematic overestimates if these lenses
are superimposed on the ``mass sheet'' of a
high-density environment \citep{hs03}.
Magnitude comparison is made by transforming
SDSS $g$- and $r$-band magnitudes
to Johnson-Morgan $V$-band using
the observed transformation of \citet{sdss2jc}.
We see that our LRG lens candidates are in general
of greater velocity dispersion than
known lenses, and in the redshift range where the two
samples overlap, the LRGs are more luminous.
The brightness of the LRGs combined with the relative
faintness of the background galaxies in our
sample suggests that any confirmed LRG lenses
would be well suited to the type of detailed
lens stellar-dynamical studies described by
\citet{tk02} and \citet{kt03}.

\section{Conclusions}

We have presented a catalog of candidate
strong galaxy-galaxy gravitational lens systems
detected spectroscopically within the Sloan Digital Sky
Survey.  These systems have known
foreground {\em and} background redshifts;
only the detailed spatial alignment of foreground
and background galaxies remains unknown.
We plan to conduct integral-field spectrographic
follow-up observations of these systems,
which will allow us to resolve the spatial
distribution of the background nebular line emission
that we have detected.
These observations should confirm a
substantial number of lenses within our candidate sample,
with many lensed galaxies at lower redshift than any other
currently known lensed extra-galactic optical sources.
Any lenses confirmed within our sample will
be of considerable interest for the study
of early-type galaxy mass distributions,
and could have implications for
lens-time-delay $H_0$ measurements
and anomalous quasar-lens flux ratios.
Our sample demonstrates the feasibility of the
emission-line-based spectroscopic lens search technique
within the SDSS and other redshift surveys,
and we plan to extend the search to higher source
redshifts in the near future.
We have also developed and applied a method for abating
the influence of night sky emission-line residuals
in the 7000--9000-\AA\ range that allows us to detect
[O~{\sc iii}] and H$\beta$ emission
over the redshift range $z \sim$ 0.4--0.8
without an excess of false-positive detections.

\acknowledgments

Funding for the creation and distribution of the SDSS Archive has been
provided by the Alfred P. Sloan Foundation, the Participating
Institutions, the National Aeronautics and Space Administration, the
National Science Foundation, the U.S. Department of Energy, the Japanese
Monbukagakusho, and the Max Planck Society. The SDSS Web site is
http://www.sdss.org/.

The SDSS is managed by the Astrophysical Research Consortium (ARC) for
the Participating Institutions. The Participating Institutions are The
University of Chicago, Fermilab, the Institute for Advanced Study, the
Japan Participation Group, The Johns Hopkins University, Los Alamos
National Laboratory, the Max-Planck-Institute for Astronomy (MPIA), the
Max-Planck-Institute for Astrophysics (MPA), New Mexico State University,
University of Pittsburgh, Princeton University, the United States Naval
Observatory, and the University of Washington.

The authors thank Paul Hewett for his constructive
referee report.
ASB thanks Paul Schechter and Hsiao-Wen Chen for
valuable discussion and consultation.

\appendix
\section{Noise Modeling}
\label{noisemodel}

If the model of a purely Gaussian noise
spectrum described by $\sigma_{\lambda}$ were
correct, then the distribution of scaled
residual specific fluxes
\begin{equation}
x_{\lambda} \equiv f^{(r)}_{\lambda} / \sigma_{\lambda}
\end{equation}
across all spectra would be Gaussian with unit variance
for all wavelengths $\lambda$.
This is unfortunately not the case in our sample.
Imperfect night-sky emission-line subtraction
and other miscellaneous effects give rise to
an excess of ``high-significance'' outliers
beyond the predictions of a Gaussian model,
leading to a deluge of false-positive
astronomical emission-line candidates
when the procedure described in \S~\ref{initial}
is applied, particularly
in the 7000--9000-\AA\ region of the spectrum
where the \oiiib\ line at redshifts
$z \approx 0.4$--0.8 appears.
The most drastic solution is simply to mask all sky-afflicted
wavelengths.  Rather than concede such
vast spectral coverage
(which would drastically reduce our survey volume), we
describe the observed distribution of
scaled residual specific
fluxes $x_{\lambda}$ within
the LRG sample with a more detailed empirical noise model.
The generally Gaussian behavior of scaled
residuals at low significance combined
with the excess of high-significance residuals is well
described by a mixture of Gaussian and Laplace
distributions,
expressed parametrically as
\begin{equation}
p(x) \, dx = \left[a \exp (-x^2 / 2 \sigma_g^2)
+ b \exp (- |x| / \sigma_e) \right] \, dx
\end{equation}
\citep[For history and applications of the Laplace
distribution, see][]{kkp01}
The parameters of this distribution are wavelength-dependent,
but we suppress this dependence in out notation.
The values of $a$ and $b$ are related by normalization:
\begin{equation}
\int_{-\infty}^{+\infty} p(x) \, dx =
\sqrt{2 \pi} \sigma_g a + 2 \sigma_e b = 1 ~~.
\end{equation}
We also fix the following relations between parameters,
based on strong correlations observed in free-parameter
fits to the distribution at each wavelength:
\begin{eqnarray}
\label{cond1}
\sigma_e &=& \sigma_g - 0.38 ~~, \\
\label{cond2}
b &=& 0.09 \times \sigma_g (a + b) ~~.
\end{eqnarray}
The result is a one-parameter noise model to fit to the
distribution of $x_{\lambda}$ across the sample
at each wavelength.
(The numerical values 0.38 and 0.09 are fixed by minimizing
the sum of binned
$\chi^2$ values for fits across all wavelengths.)
We relax conditions (\ref{cond1}) and (\ref{cond2})
and fit freely for $\sigma_e$ and $b$
at a few isolated locations in the spectrum, where
the effects of sky-subtraction residuals are especially
strong and the correlations that suggest (\ref{cond1})
and (\ref{cond2}) break down---regions near 5577~\AA,
5894~\AA, 6305~\AA, and 6366~\AA.
Additionally, some regions of some spectra are
characterized by
extreme and correlated excess variance,
so for each spectrum we convolve $|x_{\lambda}|$ capped at 5
(to limit the influence of single pixels) with a 100-pixel
boxcar filter and exclude from the noise-modeling
sample any pixels within a boxcar whose value
exceeds 1.25.

We use our fitted noise model to re-scale the reported
$\sigma_{\lambda}$ values such that the new
distribution $p(x) \, dx$ of scaled residual flux values
at each wavelength
becomes Gaussian, while preserving the position of individual
$x$-values within the cumulative distribution, then
proceed as described in \S~\ref{initial}.
Both the reported noise $\sigma_{\lambda}$ {\em and} the
measured residual flux values $f^{(r)}_{\lambda}$
contain information about the actual error in the presence
of imperfect subtraction, so it is sensible to base an
effective noise rescaling on their ratio $x_{\lambda}$
in this manner.  By fitting the noise distribution parameters
independently at each wavelength, we also capture
the localized effects of individual night-sky lines.

\section{Lensing Probabilities}
\label{lensprob}

Our strategy for assessing lensing
probabilities in our sample centers on the
construction of an approximated
probability density
$p(b) \, d b$ for the unknown impact parameter $b$
of the background galaxy in each system.
The following observed quantities are input
to the calculation:
the LRG and background redshifts,
the LRG velocity dispersion, the background
\oii\ line flux received by the
$3 \arcsec$-diameter spectroscopic fiber,
and the median seeing for the spectroscopic
plate under consideration.  We also make use
of the [O~{\sc ii}] line luminosity function (LF)
reported by \citet{hogg98}.
We adopt the same SIS model for the LRG mass
distribution as was used to obtain the $\Delta \theta$
values in Table~\ref{galtable}, and
we model the background galaxies as exponential
disks with a half-light radius of approximately
3 kpc (fixed to $0 \farcs 5$ at $z = 0.5$).

For each system, we explore a range of
impact parameters $b$ from 0 to $5 \arcsec$.
At each $b$-value,
we generate a lensed image of the model background
galaxy, then convolve it with a Gaussian
point-spread function corresponding
to the median seeing reported in the
spectroscopic plate header.  We then integrate
the image over a $3 \arcsec$-diameter
circular fiber aperture centered on the
model lens.  The result is a tabulated
function $f(b)$ giving the fraction of
the intrinsic flux received by the fiber;
that is, if the total galaxy line flux
were $S$ in the absence of
lensing and limited fiber sampling, the [O~{\sc ii}]
line flux received by the spectroscopic
fiber from a background galaxy with
a source-plane offset $b$ would be
$S_{\mathrm{fib}} = f(b) S$.
In general $f(b)$ may be greater or
less than one due to the competing effects
of lens magnification and incomplete
sampling by the fiber.

Next we adopt the [O~{\sc ii}]-emitter LF
reported by \citet{hogg98} by fitting
a Schechter function to their Figure~6.
After converting
from logarithmic units,
the number of [O~{\sc ii}] emitters per unit volume
in an interval $d L$
at line luminosity $L$ is well approximated by
\begin{equation}
\phi (L) \, d L \propto L^{\alpha}
\exp (- L / L_{\star}) \, dL ~~,
\end{equation}
with $\alpha \simeq -1.3$ and
$L_{\star} \simeq 3.4 \times 10^{42}$ ergs s$^{-1}$
(the overall
normalization is unimportant for our purposes).
We make a crude conversion from their
assumed $(\Omega_M, \Omega_{\Lambda}) = (0.3, 0)$
universe to our cosmology by scaling
their reported luminosities up by a factor
of 1.2: the ratio of squared luminosity distances
in our cosmology to theirs ranges from 1.16
at $z = 0.3$ to 1.27 at $z = 1$, and the bulk of galaxies
in their study fall within this range.
Assuming the form of the LF does not evolve,
it corresponds to an intrinsic flux function
at any redshift $z$
for the number of galaxies per unit redshift
per unit solid angle within some intrinsic
flux range $d S$ about $S$:
\begin{equation}
\psi (z, S) \, d S = N(S, z) \, S
^{\alpha}
\exp (- S / S_{\star}) \, dS ~~,
\end{equation}
with the same $\alpha$ as the LF
and $S_{\star} = L_{\star} / [4 \pi D_L^2 (z)]$,
where $D_L (z)$ is the luminosity distance to
redshift $z$.  $N(S, z)$ is a flux- and
redshift-dependent normalization, the form
of which will prove unimportant.

We can now 
derive a joint probability
density function (PDF) for the observation
of an [O~{\sc ii}]-emitting galaxy behind a given
LRG at impact parameter $b$, redshift $z$,
and with line flux
$S_{\mathrm{fib}}$ in the fiber
by setting the differential
probability proportional to the corresponding
expected differential number count
and making use of the known relationship
of $S_{\mathrm{fib}}$ to intrinsic
flux $S$ through $f(b)$:
\begin{eqnarray}
p(b, z, S_{\mathrm{fib}}) \, db \, dz \,
dS_{\mathrm{fib}} &\propto&
\psi(z, S) \, d \Omega \, dz \, d S \\
\nonumber &=& \psi[z, S_{\mathrm{fib}} / f(b)] \,
(2 \pi b \, db) \, dz \, [d S_{\mathrm{fib}} / f(b)] ~~.
\end{eqnarray}
The term $d \Omega = 2 \pi b \, db$ represents the 
solid angle in the source plane of an annulus
of radius $b$ and thickness $db$.
The {\em observed} quantities $S_{\mathrm{obs}}$
and $z_{\mathrm{obs}}$ for the system are equal to
the system's true $S_{\mathrm{fib}}$- and $z$-values
plus some observational noise that is independent
of $b$, so assuming the noise is small relative
to the scale on which the joint PDF varies,
we may reinterpret
the joint PDF as an approximate
conditional PDF on $b$
given $z_{\mathrm{obs}}$ and $S_{\mathrm{obs}}$:
\begin{equation}
p (b; z_{\mathrm{obs}}, S_{\mathrm{obs}}) \, db =
N^{\prime}(z_{\mathrm{obs}}, S_{\mathrm{obs}}) \,
b \, [f(b)]^{-1} \,
\psi[z_{\mathrm{obs}}, S_{\mathrm{obs}} / f(b)] \, db ~~.
\end{equation}
The normalization
$N^{\prime}(z_{\mathrm{obs}}, S_{\mathrm{obs}})$
need not be derived explicitly, since we
can simply compute the right-hand side without it
for the relevant range of
$b$-values and normalize afterwards.
With this PDF in hand,
we can finally assign a ``lensing probability''
to the system as the integrated
probability for impact parameters
less than the critical value
for multiple imaging in the SIS model---the
``Einstein radius'', equal to
one-half the $\Delta \theta$ value
given in Table~\ref{galtable}.
Summing this probability over
all systems gives the estimate
quoted in \S~\ref{lensornot} of
19 strong lenses out of 49 candidates.

\end{document}